\documentclass{emulateapj}
\usepackage{natbib}
\usepackage{graphicx}
\usepackage{amssymb}

\def\be{\begin{equation}}
\def\ee{\end{equation}}

\catcode`\@=11 
\def\@versim#1#2{\vcenter{\offinterlineskip
        \ialign{$\m@th#1\hfil##\hfil$\crcr#2\crcr\sim\crcr } }}

\begin{document}
\shorttitle{Outflow from hot accretion flow}
\shortauthors{Feng Yuan et al.}

\title{Numerical Simulation of Hot Accretion Flows (III): Revisiting wind properties using trajectory approach}

\author{Feng Yuan\altaffilmark{1}, Zhaoming Gan\altaffilmark{1}, Ramesh Narayan\altaffilmark{2}, Aleksander Sadowski\altaffilmark{3,4}, Defu Bu\altaffilmark{1}, and Xue-Ning Bai\altaffilmark{2,5}}

\altaffiltext{1}{Shanghai Astronomical Observatory, Chinese Academy of Sciences, 80 Nandan Road, Shanghai 200030, China; fyuan@shao.ac.cn (FY); zmgan@shao.ac.cn (ZG); dfbu@shao.ac.cn (DB)}\altaffiltext{2}{Harvard-Smithsonian Center for Astrophysics, 60 Garden Street, Cambridge, MA 02138, USA; narayan@cfa.harvard.edu (RN); xbai@cfa.harvard.edu (XNB)}\altaffiltext{3}{MIT Kavli Institute for Astrophysics and Space Rsearch; asadowsk@mit.edu}\altaffiltext{4}{NASA Einstein
Fellow}\altaffiltext{5}{NASA Hubble Fellow}

\begin{abstract}
Previous MHD simulations have shown that wind (i.e., uncollimated outflow) must exist in black hole hot
accretion flows. In this paper,  we continue our study by investigating the detailed properties of
wind, such as mass flux and poloidal speed, and the mechanism of wind production. For this aim, we make use of a three dimensional GRMHD simulation of hot accretion flows
around a Schwarzschild black hole. The simulation is designed so that the magnetic flux is not accumulated significantly around the black hole. To distinguish
real wind from turbulent outflows, we track the trajectories of the virtual Largrangian particles from
simulation data. We find two types of real outflows, i.e., a quasi-relativistic jet close to the axis and a
sub-relativistic wind subtending a much larger solid angle. We confirm that the mass flux of wind is very
significant and most of the wind originates from the surface layer of the accretion flow. The radial profile of
the wind mass flux can be described by $\dot{M}_{\rm wind}\approx\dot{M}_{\rm BH}(r/20r_s)$, with
$\dot{M}_{\rm BH}$ being the mass accretion rate at the black hole horizon and $r_s$ being the
Schwarzschild radius. The poloidal wind speed almost remains constant once they are produced, but the
flux-weighted wind speed roughly follows $v_{\rm p, wind}(r)\approx 0.25 v_k(r)$, with $v_k(r)$ being the Keplerian speed at radius $r$. The mass flux of jet is
much lower but the speed is much higher, $v_{\rm p,jet}\sim (0.3-0.4) c$. Consequently, both the energy and momentum fluxes of the
wind are much larger than those of the jet. We find that the wind is produced and accelerated primarily by
the combination of centrifugal force and magnetic pressure gradient, while the jet is mainly accelerated
by magnetic pressure gradient. Finally, we find that the wind production efficiency
$\epsilon_{\rm wind}\equiv\dot{E}_{\rm wind}/\dot{M}_{\rm BH}c^2\sim 1/1000$, in good agreement
with the value required from large-scale galaxy simulations with AGN feedback.
\end{abstract}

\keywords{galaxies: AGN - BH: feedback - BH physics: re-radiation}

\section{Introduction}

Black hole accretion models can be divided into two classes based on the temperature of the accretion
flow. One is cold accretion models such as the standard thin disk (Shakura \& Sunayev 1976;
Pringle 1981), the other is the hot accretion flow such as advection-dominated accretion flow
(Narayan \& Yi 1994, 1995; see Yuan \& Narayan 2014 for the recent review of the theory of hot accretion
flow and its applications). Hot accretion flows are believed to exist in low-luminosity active galactic nuclei,
which likely reside in majority of galaxies, and hard/quiescent states of black hole X-ray binaries.
One important question in the study of hot accretion flows in the recent years is related to winds, i.e.,
uncollimated mass outflow. On the one hand, whether winds are present or not is a fundamental question
in the dynamics of accretion, and the presence of wind help explain many observations, including the
spectrum of black hole sources (e.g., Yuan, Quataert \& Narayan 2003), emission lines from accretion
flow (e.g., Wang et al. 2013), the {\it Fermi} bubbles in the Galactic center (Mou et al. 2014), and perhaps
even some direct observations on outflow (e.g., Crenshaw et al. 2003; Crenshaw \& Kraemer 2012;
Yuan, Bu \& Wu 2012). On the other hand, AGN feedback is widely believed to play a crucial role in
galaxy formation and evolution (Fabian 2012; Kormendy \& Ho 2013), while winds produced by the AGN
accretion flow is one of the most important ingredients for such a feedback because they could interact
with the ISM in the host galaxy by exchanging momentum and depositing energy
(King 2003; Debuhr et al. 2010; Ciotti, Ostriker \& Proga 2010; Ostriker et al. 2010; Novak et al. 2011;
Gan et al. 2014). The aim of the present work is to understand the detailed wind properties
from accretion flow, which will provide important input for studies of AGN feedback.

The study of winds from hot accretion flows can be traced back to Stone, Pringle \& Begelman (1999;
see also Igumenshchev \& Abramowicz 1999, 2000). They performed the first global hydrodynamical numerical
simulation of hot accretion flow and calculated  the following time-averaged radial profiles of inflow
and outflow rates,
\begin{equation}
 \dot{M}_{\rm in}(r) = 2\pi r^2 \left\langle \int_{0}^{\pi} \rho \min(v_{r},0)
   \sin \theta d\theta \right\rangle_{t\phi},
   \label{inflowrate}
\end{equation}
\begin{equation}
 \dot{M}_{\rm out}(r) = 2\pi r^{2} \left\langle \int_{0}^{\pi} \rho \max(v_{r},0)
    \sin \theta d\theta \right\rangle_{t\phi},
\label{outflowrate}
\end{equation}
where the angle brackets represent time averages (and also average over the azimuthal angle
 $\phi$ in the case of 3D simulations). We emphasize this point because the order of doing
 time-average and the integral will make significant differences, as we will show later in this paper.
Note also that the outflow rate calculated by eq. (\ref{outflowrate}) does not necessarily represent
the mass flux of ``real outflow'', because the positive radial velocity may just come from the
turbulent motion of the accretion flow. The most important result they obtained is that the inflow
rate based on eq. (\ref{inflowrate}) follows a power-law function of radius,
\be
\dot{M}_{\rm in}(r)=\dot{M}_{\rm in}(r_{\rm out})\left(\frac{r}{r_{\rm out}}\right)^s.
\label{inflowratepowerlaw}\ee
Here $\dot{M}_{\rm in}(r_{\rm out})$ is the mass inflow rate at the outer boundary $r_{\rm out}$.
The radial dynamical range of this simulation is rather small, spanning about two orders of
magnitude in radius. But the results were later confirmed by simulations with a much larger radial
dynamical range of four orders of magnitiude (Yuan, Wu \& Bu 2012). Moreover, MHD simulations
yield very similar results that typically $s\sim 0.5-1$ (e.g., Stone \& Pringle 2001; Hawley \& Balbus 2002;
Igumenshchev et al. 2003; Pen et al. 2003; see review in Yuan, Wu \& Bu 2012).

It is exciting to note that the predicted inward decrease of accretion rate has soon been confirmed by
two observations, both are on Sgr A*. One is the detection of
radio polarization at a level of $2-9\%$ (e.g., Aitken et al. 2000; Bower et al. 2003; Marrone et al. 2007).
Such high polarization requires that the mass accretion rate close to the black hole horizon must be
within a certain range, which is two orders of magnitude lower than the Bondi rate obtained from
Chandra observations.  The other evidence is from the Chandra observation of the iron emission lines
originated from the hot accretion flow (Wang et al. 2013). The modeling to the $K\alpha$ lines indicates
a flat radial density profile near the Bondi radius, which confirms that the mass accretion rate
decreases with decreasing radius. This is because, if the mass accretion rate were a constant of radius, the density profile would be much steeper.

Two competing models have been proposed to explain the above numerical simulation result. In the
adiabatic inflow-outflow solution (ADIOS), the inward decrease of mass accretion rate is due to the
mass lost in the wind (Blandford \& Begelman 1999; 2004; Begelman 2012). In the early works of
Blandford \& Begelman (1999; 2004), the value of parameter $s$ in eq. \ref{inflowratepowerlaw} is a free
parameter. But in the more recent work of Begelman (2012), the value of $s$ is argued to be close to
unity. The other model is the convection-dominated accretion flow (CDAF) model. In this model, the
accretion flow is assumed to be convectively unstable. The inward decrease of accretion rate is
explained as more and more gas is locked in convective eddies (Narayan et al. 2000;
Quataert \& Gruzinov 2000; Abramowicz et al. 2002; Igumenshchev 2002). For a long time, it is unclear
which scenario is physical.

Three numerical simulation works have been conducted to investigate this problem (Narayan et al. 2012;
Yuan, Bu \& Wu 2012; Li, Ostriker \& Sunyaev 2013). Both Narayan et al. (2012) and
Yuan, Bu \& Wu (2012) found that the hot accretion flow is convectively stable. This indicates that the
CDAF model may not apply, leaving outflow/wind as the only possible solution. The fundamental question
is, how strong the wind is. Narayan et al. (2012) calculated the outflow rate based on eq. (\ref{outflowrate}),
except that they move the $t\phi$ average inside the integral. Although this calculation underestimates the mass flux of real outflow, as we will show later in this paper, it  eliminates contributions from turbulent
motion and produces substantially lower outflow rate than eq. \ref{outflowrate}.
In fact, only upper limit was reported since the outflow rate was found to not converge with time.
On the other hand, Yuan, Bu \& Wu (2012) systematically compared the properties of inflow and outflow,
such as angular momentum and temperature, and found that they are quite different.
They therefore concluded that systematic outflow must exist and the outflow rate must be a significant
fraction of that indicated by eq. (\ref{outflowrate}).
They argued that the rather weak outflow rate obtained in Narayan et al. (2012) is because outflow is
intrinsically instantaneous. The outflow stream can wander around in 3D space thus will be cancelled if
the time-average is performed first. Their work indicates that the mass lost via the wind is the reason for the inward decrease of the accretion rate (eq. \ref{inflowratepowerlaw}). The hydrodynamical simulations by Li, Ostriker \& Sunyaev (2013) obtained a similar conclusion as Yuan, Bu \& Wu (2012). Begelman (2012) and Gu (2014) studied why winds should exist in hot accretion flows.

The aim of the present work is two fold. First, Yuan, Bu \& Wu (2012) showed the existence of outflow
only based on some indirect arguments, so it is necessary to show the existence of wind in a more
direct way. Second, we want to quantitatively calculate the properties of wind, including the mass flux,
angular distribution, and velocity. These properties are especially important to determining how effective
the interaction between wind and ISM is. The amount of mass flux of wind is also useful to resolve the
discrepancy on the mass flux of wind between Narayan et al. (2012) and Yuan, Bu \& Wu (2012).
Sadowski et al. (2013) have also studied the properties of wind and jet under various parameters such as black hole spin. Same with Narayan et al. (2012), their calculations are based on the
time-averaged quantities thus may only give a lower limit.

We use a ``trajectory'' approach in this work for our above-mentioned aims. That is, we use the
numerical simulation data of hot accretion flow to follow the trajectories of some ``test particles''
so as to see whether the particles can really escape or simply have turbulent motions, and further
calculate the properties of the wind. The paper is organized as follows. In \S2, we will describe the
simulation data based on which we perform the analysis, and the ``trajectory'' method we use.
The results are presented in \S3. \S4 is devoted to the analysis of the acceleration mechanism of
wind. We then summarize and conclude in \S5.

\section{Simulation data and the trajectory method}

\subsection{Simulation data of hot accretion flow models}
\label{simulationdata}

We have considered two different simulations to perform the ``trajectory'' analysis.
One is the two dimensional MHD simulation described in Yuan, Bu \& Wu (2012), and the other is
the three dimensional GRMHD simulation described in Narayan et al. (2012).
In the former, the initial condition is a rotating torus with constant specific angular momentum. The
density maximum is located at $r=100r_s$, where $r_s\equiv 2r_g\equiv 2GM/c^2$. The initial
magnetic field is poloidal, a single set of loops confined to the interior of the torus, and the loops are parallel to the density loops.
The simulation is performed using the ZEUS code. The readers are referred to Yuan, Bu \& Wu (2012)
for details. In the second simulation, the initial condition is again a rotating torus, but the details of the
torus are different. It has inner and outer edges at $r_{\rm in}=10 r_g$ and $r_{\rm out}=1000r_g$,
respectively. The simulation domains ranges from close to the black hole to $\sim 10^5 r_g$. The initial magnetic field is again purely poloidal. Different from the first simulation, the magnetic field is broken
into eight poloidal loops of alternating polarity. Each loop carries the same amount of magnetic flux, so
that the black hole is unable to acquire a large net flux over the course of the simulation. The simulation
is performed using the HARM code. It is run for a time of $2\times10^5GM/c^3$ and achieves inflow
equilibrium (i.e., accretion has reached a steady state) out to a radius $\sim 90r_g$. The readers are referred to Narayan et al. (2012) for details.
By using the trajectory approach to study the wind properties using data from both simulations, we find
that the main results are very similar. Therefore, in this paper, we choose to only focus on the latter
simulation (3D GRMHD). Throughout this paper, we use the spherical coordinate. The physical quantities in the present paper, if not specified, are in units of $G=M=c=1$.

\subsection{Trajectory method}

Trajectory is related to the Lagrangian description of fluid, obtained by following the motion of fluid
elements at consecutive times. Streamline is associated with the Euler description of fluid, obtained
by connecting the velocity vectors of adjacent fluid elements at a given time. Trajectory is only
equivalent to the streamline for strictly steady motion, which is not the case for accretion flow since
it is always turbulent. Streamlines are easy to obtain and are widely used in the literature. On the
other hand, obtaining Lagrangian trajectories is much more time-consuming than streamlines, but
they loyally reflects the motion of fluid elements. For our purpose, we should consider trajectories
rather than streamlines.

To get the trajectory, we first choose a set of ``test particles'' in the simulation domain within the outmost radius where inflow equilibrium is achieved, $\sim 90r_g$,  at a given snapshot at time $t$ of the simulation. They are not real particles, but a collection of spatial
coordinates as the starting point for the trajectory calculation. With their velocities interpolated
from simulation data, we can then obtain their location at time $t+\delta t$. This process is then
continued until the end of the simulation or when test particles leave the simulation domain.
We use the ``VISIT'' software to perform particle trajectory calculation, which can do interpolation
with a controlled precision. Obviously, to obtain robust particle trajectories, the time step of the
simulation data output $\delta t$ must be sufficiently small. This timescale depends on how fast the velocity of the particles change during their motion, which is the Keplerian timescale. So the Keplerian timescale must be properly time-resolved.
We have tested in our calculations using different time resolutions and compared the corresponding
trajectory to see whether the results converge. If not, we shorten the time step of the simulation
data output. The time step we actually use in obtaining the trajectory is roughly
the Keplerian timescale at $r\approx 6r_g$, which is much shorter than the Keplerian timescale of
most of the region of the accretion flow. Taking 100 particles at various radius as example, we have tested two time steps, with one being the Keplerian timescale at $6r_g$ and another the Keplerian timescale at $4r_g$. We found that the results such as the particle trajectories are largely indistinguishable.

\section{Results}

\begin{figure*}
\vspace{0cm}\hspace{-0.0cm}\epsscale{0.6} \plotone{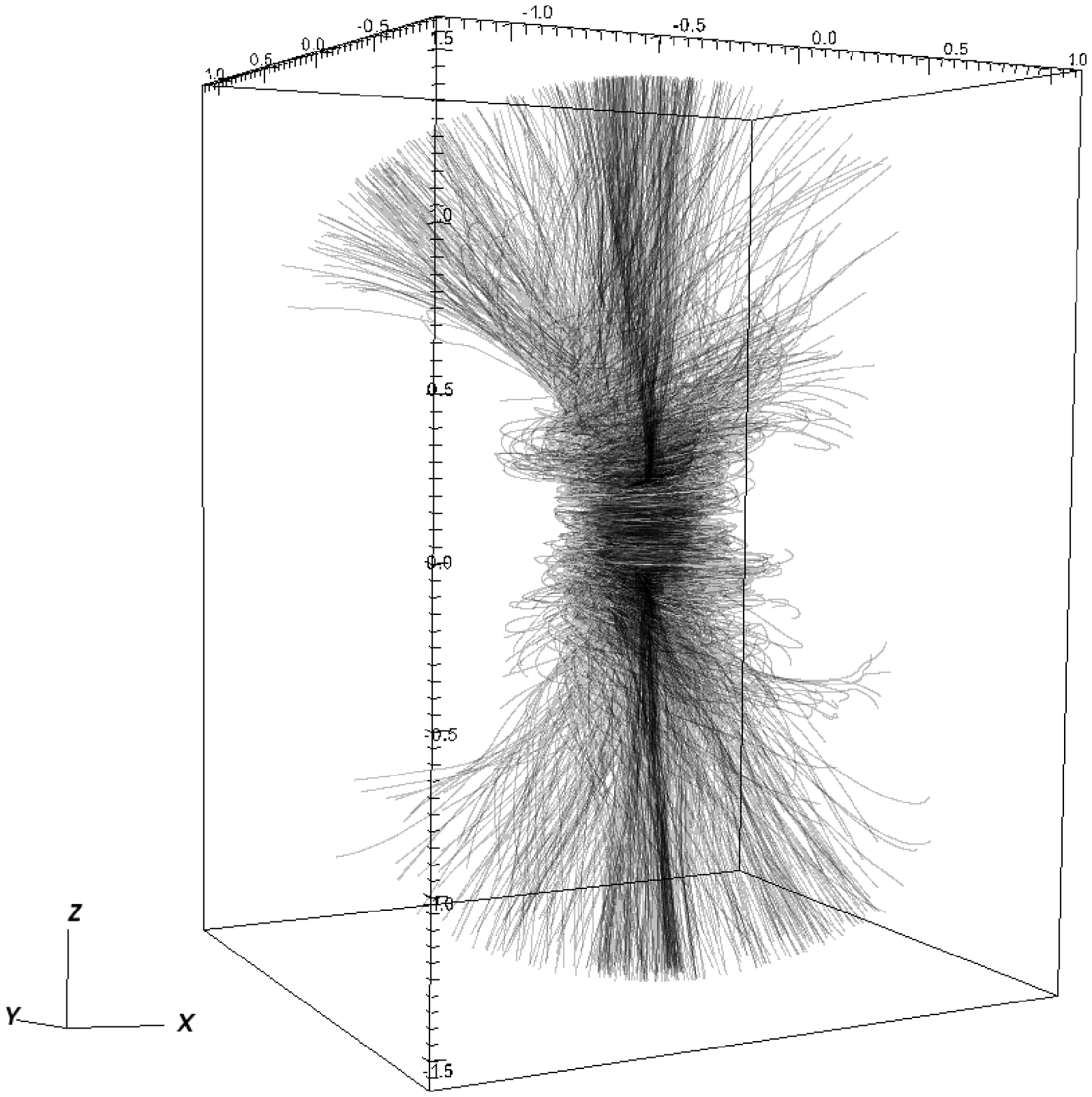}\vspace{0.0cm}\hspace{-0.15cm} \epsscale{0.35}
\plotone{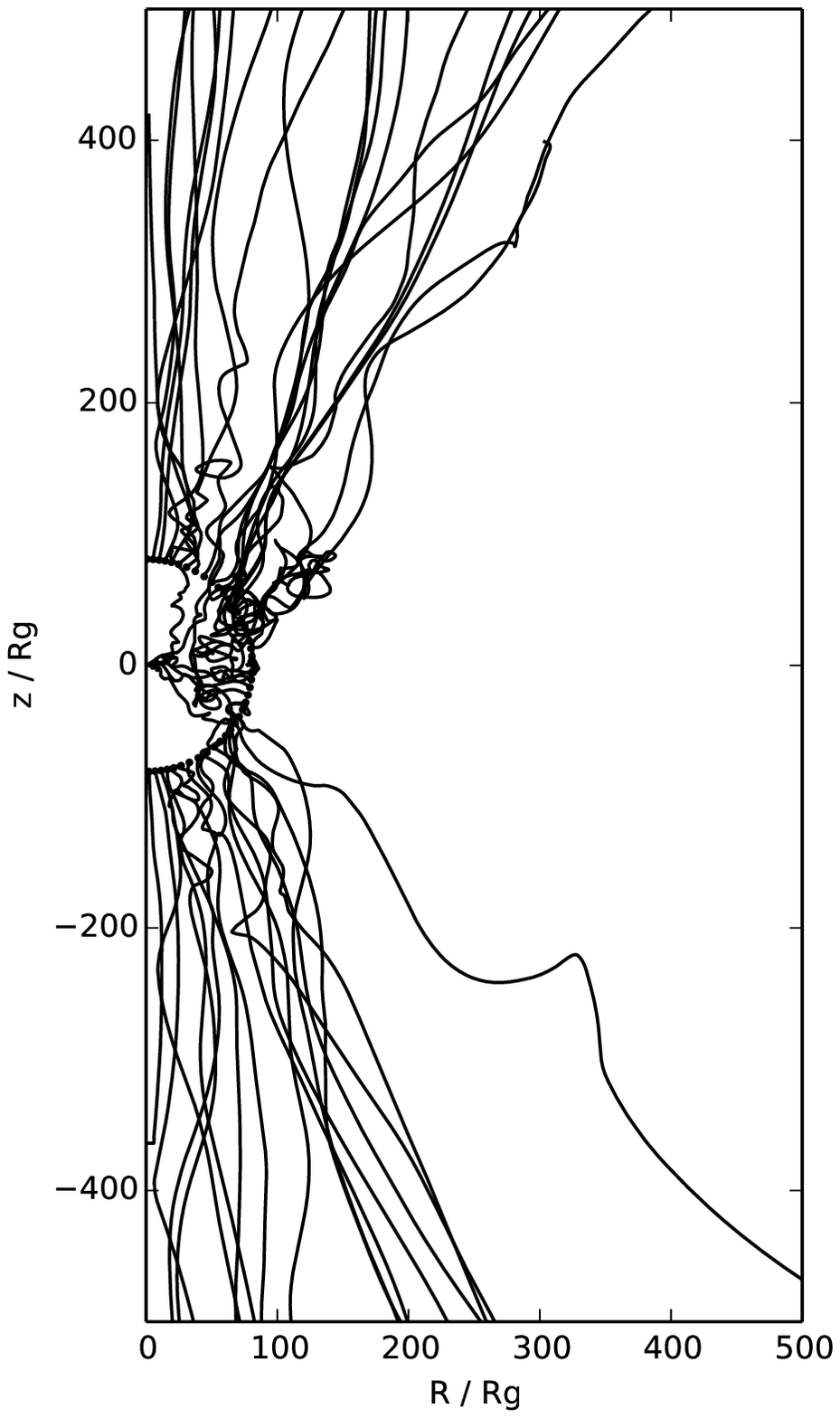}
\vspace{0.cm}
\caption{Lagrangian trajectories of the ``test particles'' originating from $r=80r_g$ (the black circle in
the right panel) in the three-dimensional space ({\it left}) and two-dimensional ($r-\theta$) plane
({\it right}). Real outflows are evident in the coronal region.
The inflow concentrates within the main disk body around the equatorial plane and their motion is
more turbulent.}\label{Fig:trajectory}
\end{figure*}

Before describing our results, we first define some terminologies.
We use ``outflow'' to describe any flow with a positive radial velocity $v_r$, i.e., flowing outward.
This includes both ``turbulent outflow'' and ``real outflow''.
The difference between them is that in the former case the test particle will eventually return and join the
accretion flow after flowing outward for some distance, while in the latter case the test particle continues
to flow outward and eventually escapes the outer boundary of the simulation
domain\footnote{In some cases, e.g., if the initial radius of test particles are small, some particles could
not escape beyond the outer boundary by the end of the simulation. On the other hand, they keep moving
radially outward without any signs of return (the poloidal speed does not decrease outward, refer to
Fig. \ref{Fig:poloidalspeed}), we regard them as representing real outflow.}. ``Real outflow'' consists of two
components, i.e., disk jet and wind. Here, ``disk jet'' is different from the Blandford-Znajek jet in several
aspects, as we have summarized in Yuan \& Narayan (2014). Very briefly, ``disk jet'' originates from the
innermost region of the accretion flow. It is quasi-relativistic and matter-dominated. The Blandford-Znajek
jet is powered by the spin of the black hole. It is relativistic and Poynting flux-dominated. In the present paper, we do not have a Blandford-Znajek jet since the simulation data we use is for a Schwarzschild black hole. As we will
describe in detail later, we find that the disk jet is confined in a region of $\theta\la 15^{\circ}$ away from
the polar axis while the wind is located between the jet boundary and the surface of the accretion flow.
The velocity of the jet is much higher than that of the wind.

Note that the definition of wind we adopt here is different from that adopted in some literature (e.g., Narayan et al. 2012; Sadowski et al. 2013), where they require that the Bernoulli parameter of  wind must satisfy $Be>0$.
The Bernoulli parameter of our wind can be of any value, negative or positive, at any radius. Our argument is that for non-steady accretion flow, $Be$ is not constant along trajectories, but usually increases outward (refer to Fig. \ref{Fig:bernoulli}). 
This means that even though $Be<0$ at a certain radius, the wind particles can still escape to infinity. In fact, as we will describe later, we find that no matter what value $Be$ is, the poloidal speed of wind does not decrease when they propagate outward until a radius within which turbulence is developed (refer to Fig. \ref{Fig:poloidalspeed} and relevant discussions later). In addition, technically the value of $Be$ of wind should depend on the initial condition of the simulation because of energy conservation. In many simulations including the current one, the initial condition is a bound torus thus $Be$ is negative. But in reality, the accretion flow comes from much further away so $Be$ is more likely positive. This implies that the value of $Be$ obtained in simulation should be regarded as an lower limit.

\subsection{Overall result: confirmation of the wind}
\label{overallresult}

\begin{figure}\vspace{-0.cm}
\epsscale{1.}\plotone{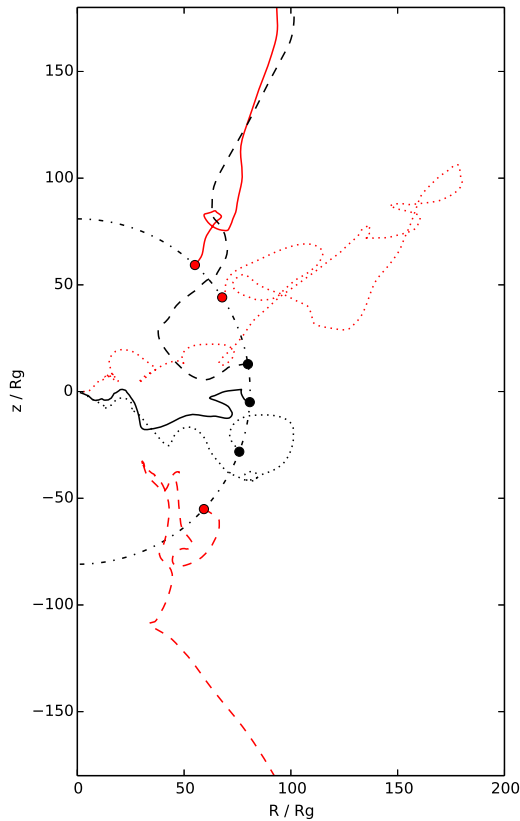}\epsscale{1.}
\vspace{-1.5cm}
\caption{The various types of trajectory of ``test particles'' in the accretion flow. Red lines denote outflow
while black ones for inflow. See section 3.2 for details. }
\label{Fig:sixtrajectory}
\end{figure}

Figure \ref{Fig:trajectory} shows the trajectories of sample test particles originated from locations
$(r, \phi)=(80r_g,0)$ and $\theta= 0 - \pi$. The left and right panels show the trajectories at the three and two ($r-\theta$ plane) dimensional space, respectively.
The results are similar for test particles originating from different radii.

From the figure we clearly see that winds are evident. They are largely located in the polar region,
i.e., $\theta\sim 0-\pi/4$ and $\theta=3/4\pi-\pi$, and are symmetric with respect to the equatorial plane.
The trajectories of many test particles in this region are almost straight lines in the poloidal plane and $\theta\sim$ const., indicating that turbulence is weak. There are also many other more turbulent trajectories
with varying $\theta$ values during outflowing motion. 
The specific angular momentum of wind particles is found to be larger than that of the inflow, consistent with Yuan, Bu \& Wu (2012). The wind region almost overlaps with the usual ``coronal'' region above the main body of accretion flow
(refer to Fig. 4 in Yuan \& Narayan 2014 for the structure of accretion flow), and the boundary of the wind
region is about the surface of the accretion flow defined by the density scale height. In other words, the
disk corona is outflowing. The main disk body, i.e., $\theta=\pi/4-3/4\pi$, is the ``inflow'' region. Most of
the test particles originating from this region move inward, and their motion is turbulent. This turbulent
motion is due to the magneto-rotational instability (MRI; Balbus \& Hawley 1998), as usual. Interestingly,
we also find that some test particles originated from this region first move vertically toward the coronal
region, and then escape outward radially as disk wind. Such vertical motion is present in almost any
radius and is perhaps an indicator of the magnetic buoyancy. This supplies new gas from the disk body to the corona/wind.
Overall, the structure of the accretion flow is that the inflow region corresponds to the main disk body,
while the wind region corresponds to the disk corona. This picture is consistent with
Sadowski et al. (2013; see their Fig. 16).

\subsection{The mass flux of outflow and inflow}


\begin{figure*}
\hspace{-0.35cm}\epsscale{0.6} \plotone{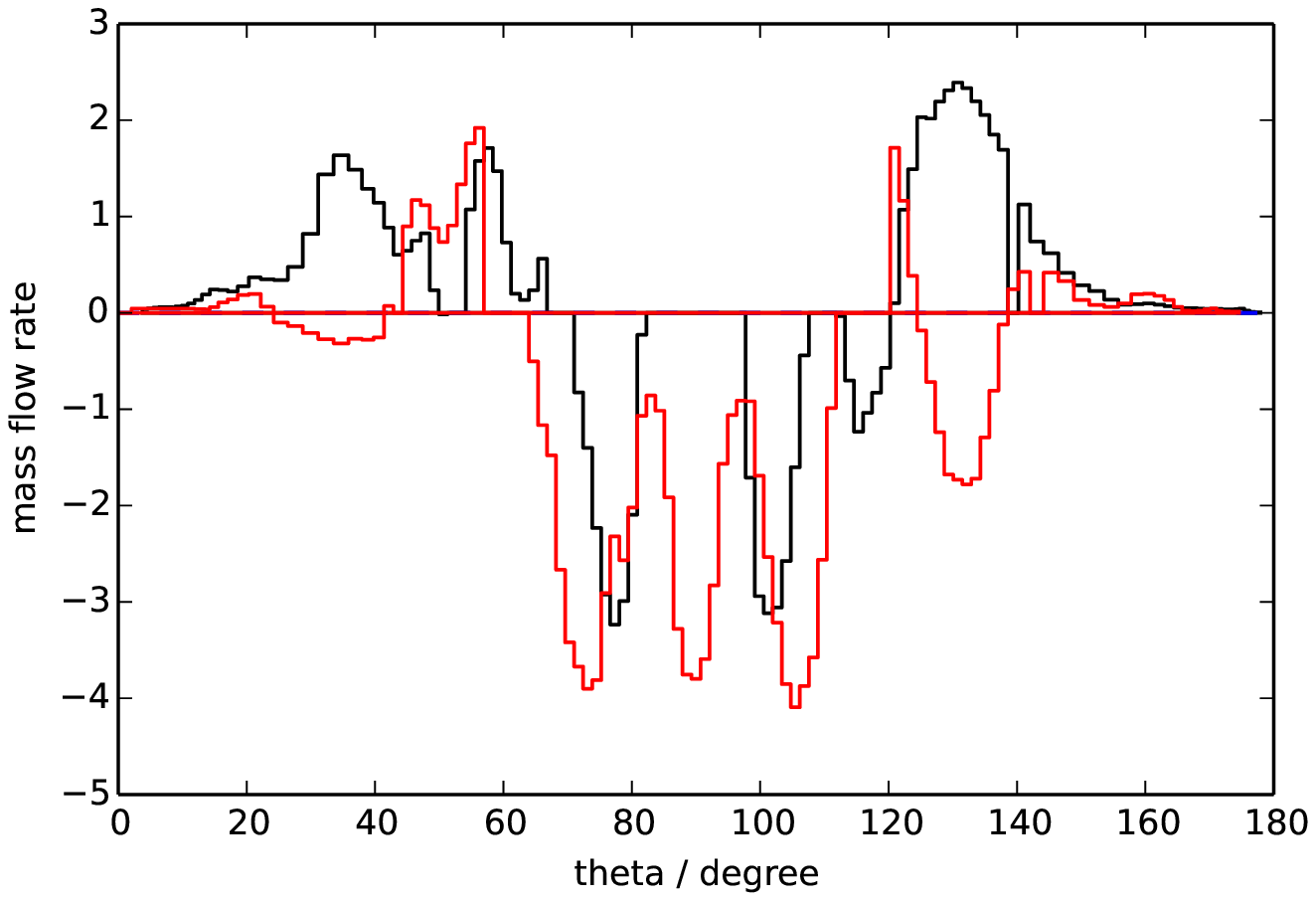}\hspace{-0.5cm} \epsscale{0.6}
\plotone{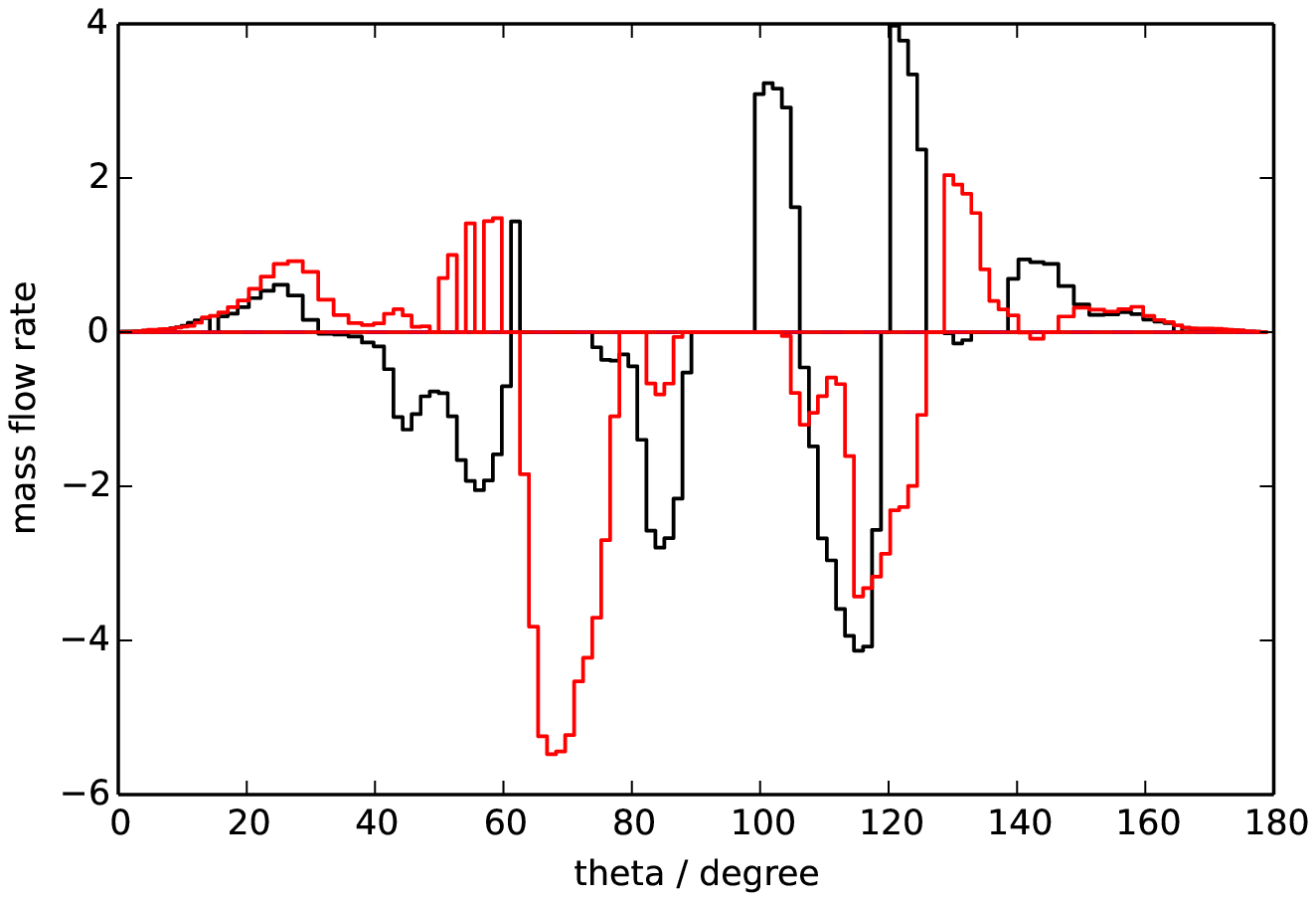}
\vspace{0.0cm}
\caption{The mass flux per unit $\theta$ and $\phi$ at $r=80r_g$ as a function of $\theta$. {\it Left}: Black
and red lines are for $t=104000M$ and $120000M$, respectively, at fixed $\phi=0$; {\it Right}: Black and
red liens are for $\phi=5\pi/4$ and $3\pi/2$, respectively, but fixed time $t=100000M$.
Positive value corresponds to real outflow while the negative value corresponds to total inflow
(i.e., including turbulent inflow). }
\label{Fig:angularrate}
\end{figure*}

\begin{figure}
\hspace{-0.35cm}\epsscale{1.1} \plotone{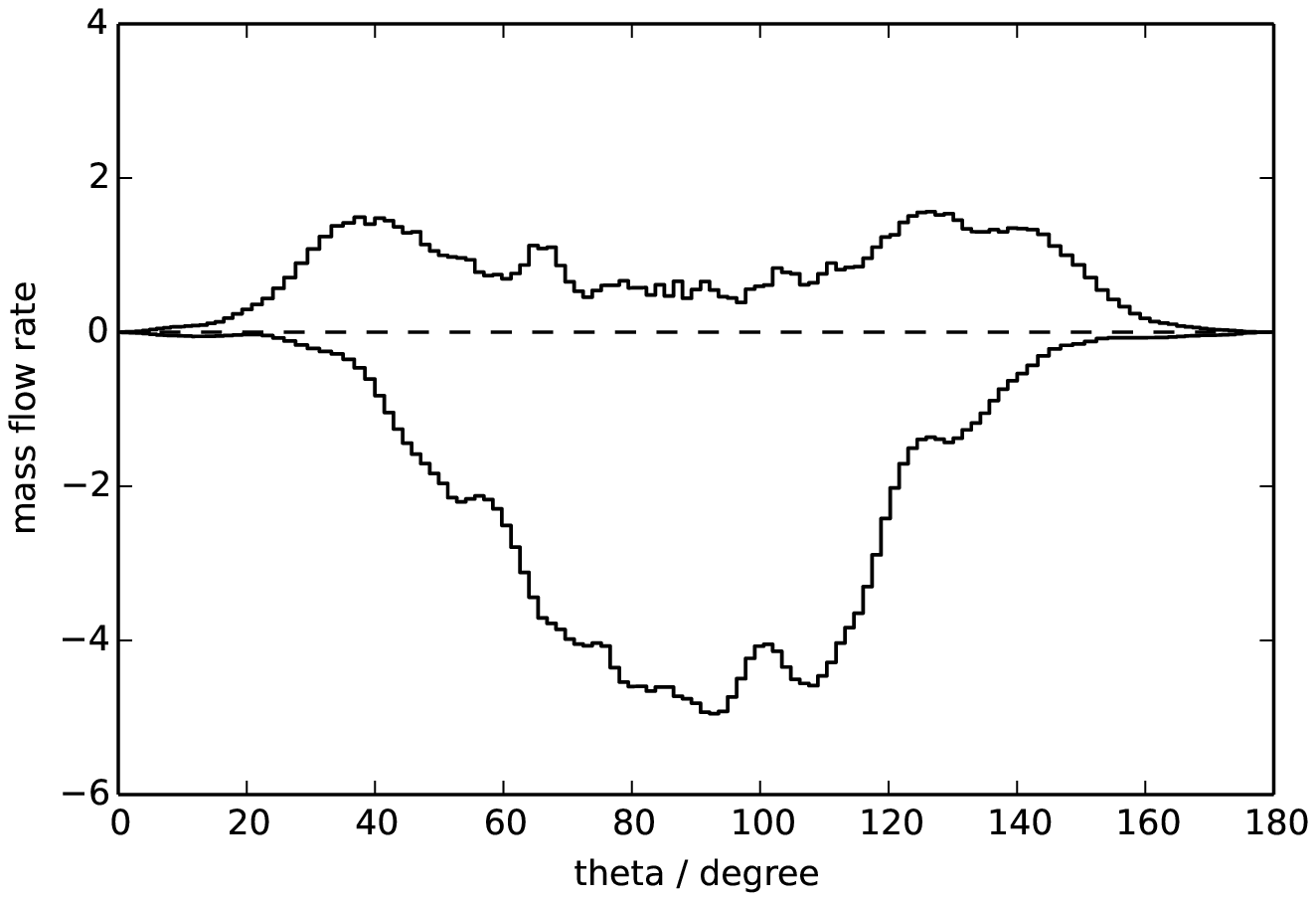}
\vspace{0.5cm}
\caption{The mass flux per unit $\theta$ but integrated over all $\phi$ and averaged from time $t=100000-120000M$ as a function of $\theta$ at radius
$r=80 r_g$. The positive and negative values are for the real outflow and total
inflow, respectively. Note that significant mass flux of wind are also produced from the disk body, i.e,
around $\theta\sim 100^{\circ}$. These wind particles first move vertically toward the disk surface and
then escape outward from there.}
\label{Fig:massrate40}
\end{figure}

To calculate the mass flux of the inflow and outflow, and to analyze the wind properties, we
distinguish between various types of particle trajectories.
In Figure \ref{Fig:sixtrajectory}, we show characteristic types of characteristic particle trajectories originated
from radius $r$. Distinguishing them is crucial for calculating the mass fluxes of the inflow and outflow
correctly, as well as analyzing the wind properties.
\begin{itemize}
\item The red and black lines represent outflows and inflows, respectively.
\item The red solid line represents a real outflow, where the particle keeps moving outward and
never cross the radius $r$ during its motion.
\item The red dashed and red dotted lines represent turbulent outflows, where the particles first move
outward, but will later return and cross the radius $r$ during their motion. Although the particles
eventually move outward and inward, they both belong to the ``turbulent outflow'' category when
we calculate the mass flux.
\item It is similar for the three black lines except that they all represent inflow. The solid line represents
real inflow, while both the dashed and dotted lines represent turbulent inflow.
\end{itemize}

To calculate the mass flux (or mass flow rate) of the wind $\dot{M}_{\rm wind}(r)$ at a given time, we
first choose test particles initially distributed at fixed radius $r$ with different $\theta$ and $\phi$ and
obtain their trajectories.  The ``real outflow rate'' is then calculated by summing up the corresponding
mass flux  carried by test particles whose trajectories belong to the real outflow (the red solid line in
Fig. \ref{Fig:sixtrajectory})
\begin{equation}
\dot{M}_{\rm wind}(r)=\sum_i \rho_i(r) v_{r,i}(r)r^2 \sin(\theta) \delta \theta_i \delta \phi_i.
\end{equation}
Here $\rho_i(r)$ and $v_{r,i}$ are the mass density and radial velocity at the location where test
particle ``i'' originates, $\delta \theta_i$ and $\delta \phi_i$ are the ranges of $\theta$ and $\phi$ the
particle occupy. We can obtain the ``real inflow rate'' with a similar approach. In this case, we  sum
up the corresponding mass flux of test particles whose trajectories are analogous of the black solid
line in Fig. \ref{Fig:sixtrajectory}. The mass flux corresponding to the dotted and dashed red lines
are ``turbulent outflow rate'', while that corresponding to the dashed and dotted black lines are are
``turbulent inflow rate''.

\begin{figure*}
\hspace{-0.0cm}\epsscale{1.1} \plotone{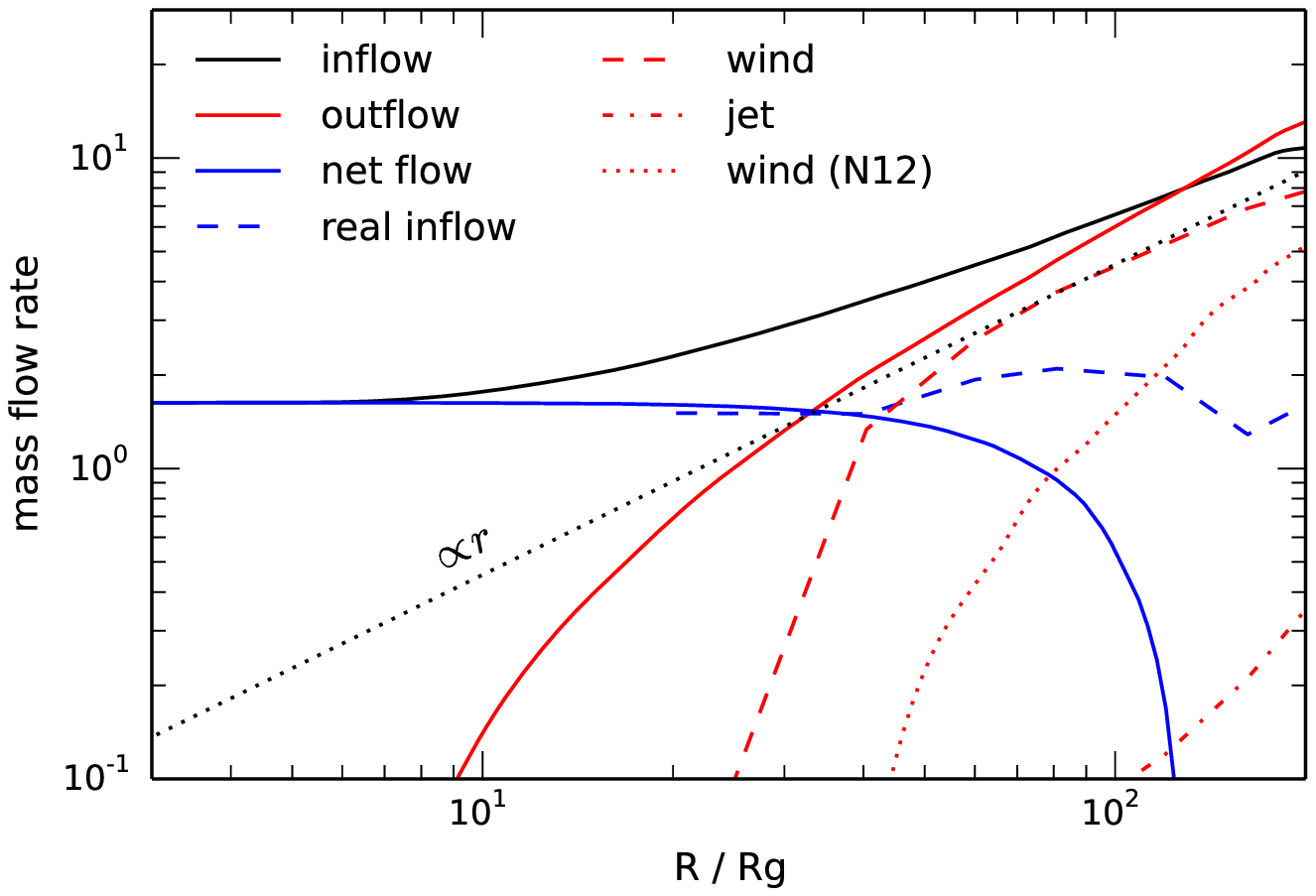}
\vspace{0.2cm}
\caption{Various mass flow rates as a function of radius. Black and red solid lines show the total inflow
and outflow rates calculated following eqs. (\ref{inflowrate}-\ref{outflowrate}), while the blue solid line
denotes their difference. The red and blue dashed lines denote the mass flux of the real outflow and
real inflow calculated at $t=100000M$, respectively. The red dot-dashed line shows the mass flux of
the disk jet. For comparison, we also show by the red dotted line the mass flux of real outflow calculated following the method in Narayan et al. (2012).}\label{Fig:massrate}
\end{figure*}

Fig. \ref{Fig:angularrate} shows the mass flow rate  at $r=80 r_g$ per unit $\theta$ and $\phi$ as a function of $\theta$.  The left panel shows the mass flux at two different times but the same $\phi$; while the right panel shows the mass flux at two different $\phi$ but the same time. Positive value is for real outflow while negative value is for total inflow, i.e., including both the ``real inflow'' and ``turbulent inflow''. By comparing the black and red lines in the left and right panels, we see that the specific values of $\theta$ at which real outflows reside change with $\phi$ and time. In other words, for a fixed $\theta$, the flow can be inflow or outflow for different time $t$ and $\phi$\footnote{The values of time in the left panel and the values of $\phi$ in the right panel are chosen so that the change of inflow and outflow for a fixed $\theta$ is significant.}. This result indicates that if we move the $t\phi$ average inside the integrals of eq. \ref{outflowrate}, i.e., to integrate max($\langle \rho v_r\rangle_{t\phi}$,0) as in Narayan et al. (2012), significant mass flux of real outflow will be cancelled and we will substantially underestimate the mass flux of the real outflow. This is the main reason why Narayan et al. (2012) reported much weaker real outflow than Yuan, Bu \& Wu (2012). This also explains why Narayan et al. (2012) reported that a the mass flux of the real outflow does not converge with time since more mass flux will be cancelled if the period of integration time is longer. Fig. \ref{Fig:massrate40} shows the mass flow rate integrated over all $\phi$ and averaged from time $t=100000-120000M$ at radius $r=80r_g$ as a function of $\theta$. Same with Fig. \ref{Fig:angularrate}, positive and negative values are for real outflow and total inflow, respectively.

From Figs. \ref{Fig:angularrate} \& \ref{Fig:massrate40} we see that consistent with the qualitative result
shown in Fig. \ref{Fig:trajectory}, the inflow primarily takes place in the main disk body while most of
mass flux of the real outflow occurs in the range of $\theta=30^{\circ}-60^{\circ}$. But as we have
pointed out in \S\ref{overallresult}, for some values of $\phi$ at any radius, some real outflow also exists
in the main disk body. This can be seen from Fig. \ref{Fig:massrate40} which shows that some
significant mass flux of real outflow even close to the equatorial plane, i.e., $\theta\sim 100^{\circ}$. We hardly see these wind in
Fig. \ref{Fig:trajectory} because  these wind particles first quickly move vertically to the surface of the accretion flow before moving outward as outflowing wind.

The next important questions are, what is the mass flux of real outflow?  How is it compared with the total outflow rate calculated by eq. (\ref{outflowrate})?
By integrating the wind mass flux shown in Fig. \ref{Fig:massrate40} over $\theta$ but without doing time average, we can obtain the total real outflow rate at a given time and radius. Then we can obtain the radial profile of mass flux
of real outflow. The red dashed line in Fig. \ref{Fig:massrate} shows the result at $t=100000M$.\footnote{We want to point out a caveat here. As we have stated in \S\ref{simulationdata}, the inflow equilibrium is reached only up to $\sim 90r_g$ in our simulation, but the x-axis of the figure extends to $200r_g$.}
We find that the radial profile of the mass flux of wind can be well described by
\begin{equation}
\dot{M}_{\rm wind}(r)\approx\dot{M}_{\rm BH}\left(\frac{r}{40 r_g}\right)^s\equiv\dot{M}_{\rm BH}\left(\frac{r}{20 r_s}\right)^s, {\rm } ~s\approx 1,
\label{windflux}
\end{equation}
where $\dot{M}_{\rm BH}$ is the mass accretion rate at the black hole horizon. So the mass flux of the wind at $40r_g$ is equal to the mass accretion rate to the black hole $\dot{M}_{\rm BH}$. Such a power-law distribution is likely valid up to the outer boundary of the accretion flow, and then quickly decreases beyond the outer boundary (Yuan et al. in preparation). For comparison, we have also shown in the figure by the red and black solid lines the total outflow and total inflow rates calculated by eqs. (\ref{outflowrate}) and (\ref{inflowrate}), and the mass flux of real outflow calculated following the method in Narayan et al. (2012) by the red dotted line. We can see that the mass flux calculated by eq. (\ref{outflowrate}) is equal to $\dot{M}_{\rm BH}$ at $30r_g$; while that calculated by the Narayan et al. (2012) method is much weaker, equal to $\dot{M}_{\rm BH}$ at $100 r_g$. Sadowski et al. (2013) adopted the same method as
Narayan et al. (2012) to study the mass flux of real outflow  but could not obtain a radial profile, because they
found that the mass flux of the real outflow is too weak thus difficult to fit using any formula. We also note that the power-law index in eq. (\ref{windflux}) is in good agreement with that obtained from the analytical study by Begelman (2012).

Also shown in the figure by the red dot-dashed line is the mass flux of disk jet. The mass flux of disk
jet can be described by
\begin{equation}
\dot{M}_{\rm jet}(r)=\frac{1}{20}\dot{M}_{\rm wind}(r).
\end{equation}
The mass flux increases with radius, indicating that the jet is gradually supplied by matter from the wind.


\subsection{The poloidal speed}
\label{poloidalspeed}

\begin{figure*}
\epsscale{0.5}\plotone{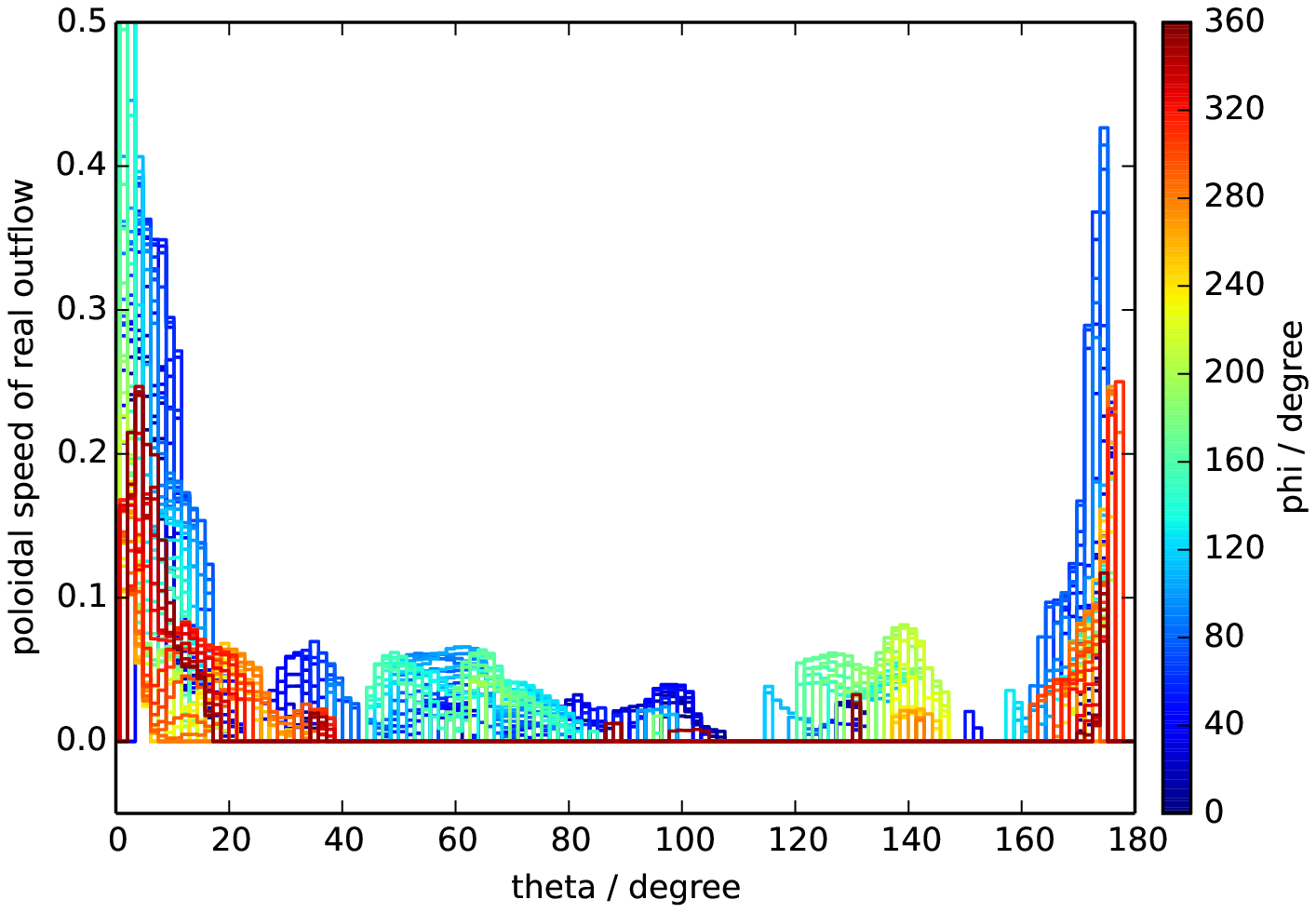}\epsscale{0.5}
\plotone{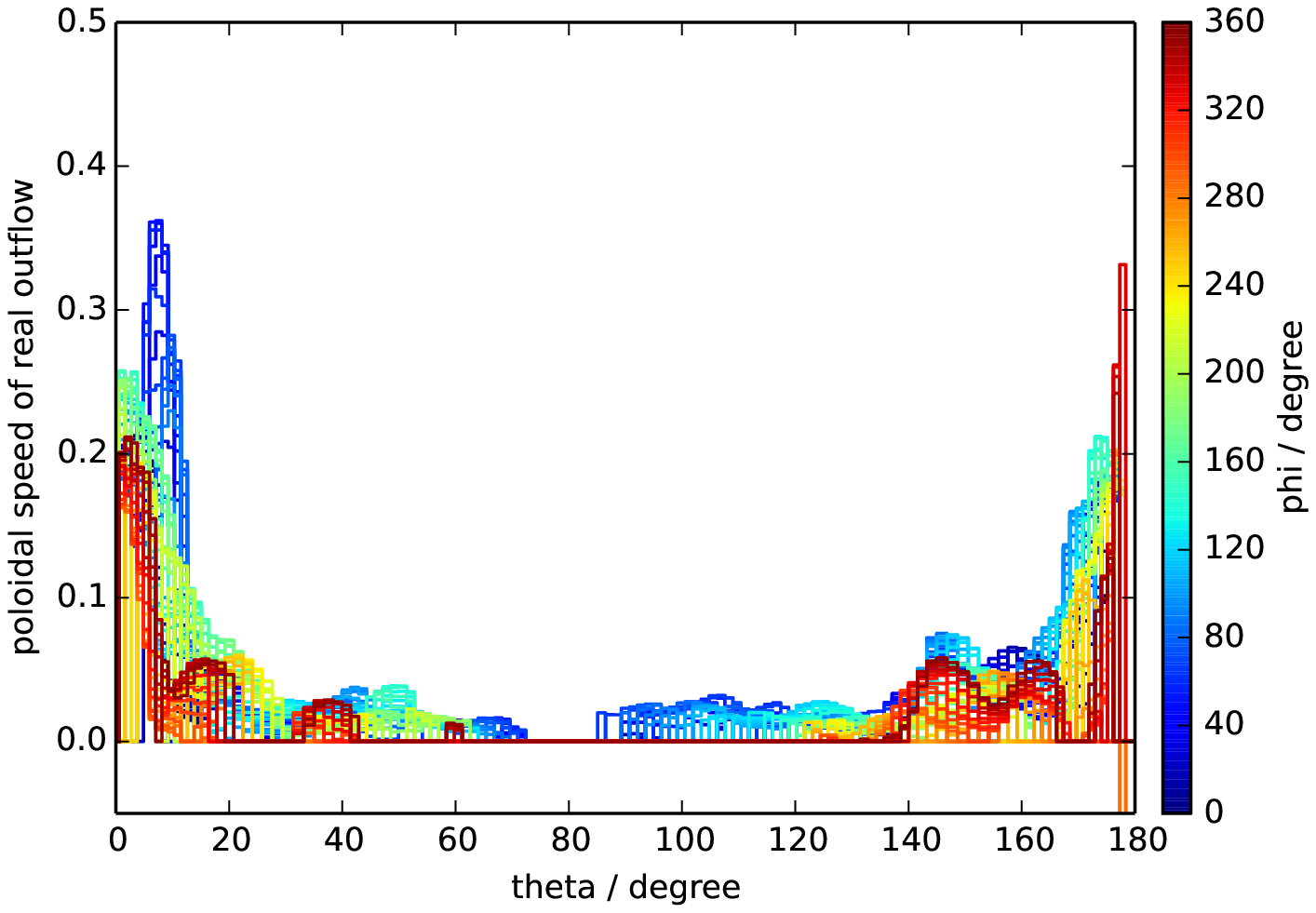}\epsscale{0.5}
\plotone{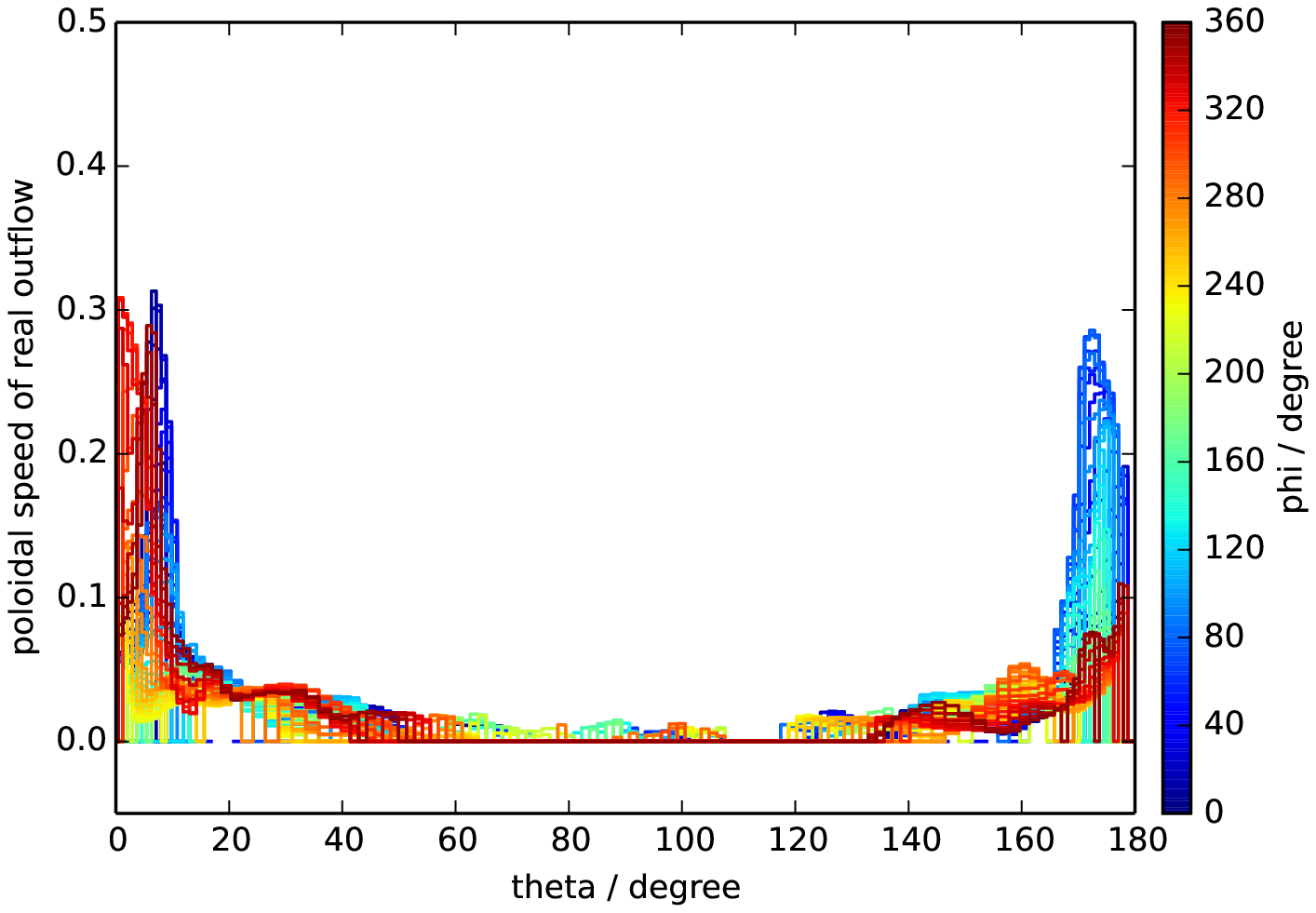}\epsscale{0.5}
\vspace{0.5cm}
\caption{The poloidal speed in unit of speed of light of real outflow as a function of $\theta$ for $r=40 r_g$ (top left), $80r_g$ (top right), $160r_g$ (bottom) and various $\phi$ at $t=100000M$. The values of $\phi$ are denoted by the color of the lines. We can see that close to the axis, $\theta\la 10^{\circ}$ and $\theta\ga 170^{\circ}$, the poloidal speed is much larger than in other regions. We identify this part of outflow jet. }
\label{Fig:poloidalspeedtheta}
\end{figure*}

\begin{figure}
\epsscale{1.}\plotone{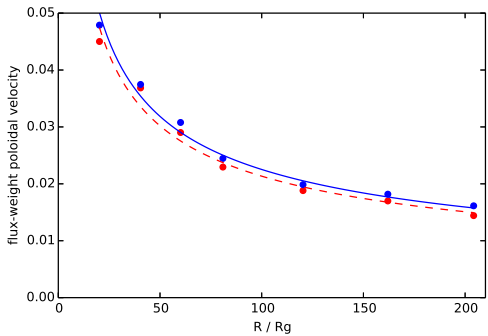}\epsscale{1.}
\vspace{0.cm}
\caption{The mass flux-weighted poloidal speed in unit of speed of light at $t=100000M$. The blue and red lines correspond to total outflow (i.e, averaged for all $\theta$) and wind (i.e., averaged over $15^{\circ}\la \theta \la 165^{\circ}$). }
\label{Fig:poloidalspeedradius}
\end{figure}

\begin{figure*}
\epsscale{0.5}\plotone{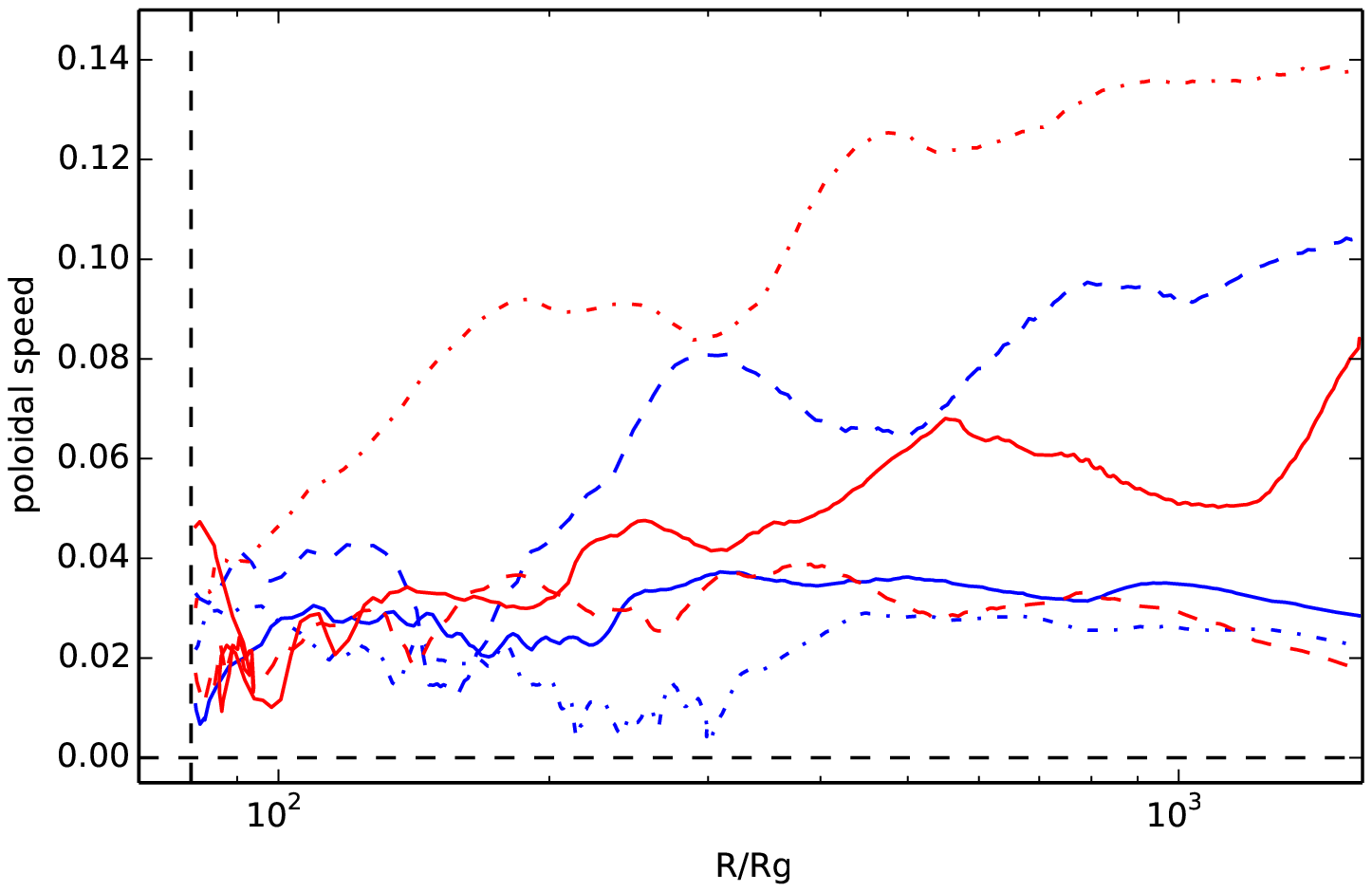}
\epsscale{0.5}\plotone{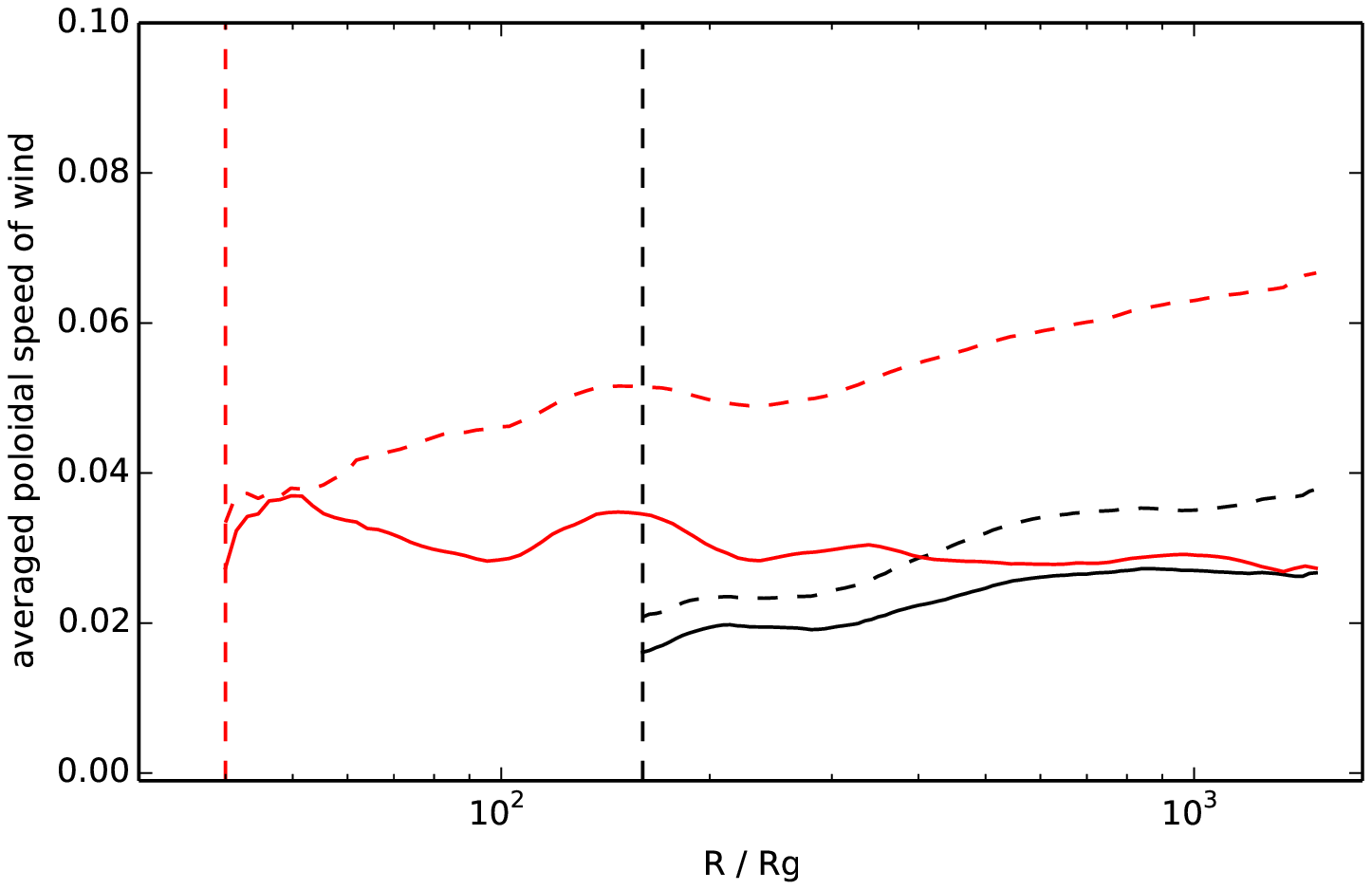}
\vspace{0.0cm}
\caption{{\it Left:} Evolution of the poloidal velocity of real outflows along test particle trajectories
(originated from $r=80r_g$). The three red lines have {\it initial} $Be>0$, while the three blue lines have
{\it initial} $Be<0$. The exact values of $Be$ along the trajectories can be found from Fig. \ref{Fig:bernoulli}.
From top to bottom, the red dot-dashed, blue dashed, and red solid lines correspond to particles originated
from $\theta\sim 10^{\circ}, 20^{\circ}, 30^{\circ}$, respectively. The red dashed, blue solid, and blue
dot-dashed lines correspond to $\theta\sim40^{\circ}\la \theta \la 50^{\circ}$.
{\it Right:} Spatially averaged
poloidal velocity of real outflows originated from 40$r_g$ (red lines) and
160 $r_g$ (black lines) along their trajectories. The dashed lines are calculated from wind particles
originated from the region of $15^{\circ}\la \theta \la 165^{\circ}$, while the solid lines are calculated from
wind particles always satisfying $15^{\circ}\la \theta \la 165^{\circ}$ in their trajectories.}\label{Fig:poloidalspeed}
\end{figure*}

In this subsection, we consider the evolution of poloidal velocity in the wind. This is the dominant
component of the wind velocity at large disk radii once magnetic field becomes subdominant.
Fig. \ref{Fig:poloidalspeedtheta} shows the poloidal speed of the real outflow as a function of $\theta$ for
various $\phi$ at three different radii, $r=40, 80$ and $160 r_g$, and $t=100000M$. To obtain these three
plots, we first choose some test particles at vavious $\theta$ and $\phi$ at the three radii and obtain their
trajectories. We then select those particles corresponding to real outflow and obtain their poloidal speed.
We can see that the poloidal speed as a function of $\theta$ has a sharp jump at $\theta\sim 15^{\circ}$
away from the rotation axis. The poloidal speed of outflow close to the axis is $\ga 0.3c$, much larger than
those away from the axis, which is $\la 0.05c$. We thus can naturally identify the real outflow within
$\theta\sim 15^{\circ}$ to the axis as ``disk jet'', while the real outflow out of this range to be wind. Note
that the simulation data we use is for a non-rotating black hole, so the presence of the disk jet is irrelevant
to black hole spin, although the spin of the black hole may strengthen the disk jet (e.g., Sadowski et al. 2013).
The disk-jet originates from the inner disk region and are powered by the rotation energy of the accretion
flow. This is different from the Blandford-Znajek jet originated from the black hole horizon
(Blandford \& Znajek 1976), which is powered by the black hole spin energy. Other differences between
the two types of jet include that the disk jet is sub-relativistic and matter-dominated, while the
Blandford-Znajek jet is relativistic and Poynting-flux dominated (see Yuan \& Narayan 2014 for a summary).

By comparing the three plots in Fig. \ref{Fig:poloidalspeedtheta}, we see that the poloidal wind velocity
seems to decrease with increasing radius. To quantify, we calculate the mass flux-weighted
poloidal speed of real outflow as a function of radius.
We distinguish the two types of the real outflow, i.e., wind and disk jet, in our calculations. For this purpose,
when calculating the poloidal wind velocity, the integration is only over the range of
$15^{\circ}\la \theta \la 165^{\circ}$; while for the entire outflow, we integrate over all $\theta$. The
results are shown by blue (total outflow) and red (wind only) dots in Fig. \ref{Fig:poloidalspeedradius}. The
poloidal velocity of the total outflow is slightly higher than that of wind, as expected. The lines are fitting
functions, approximately given by
\begin{equation}
v_{\rm p,wind}(r)\approx 0.21 v_k(r).
\label{poloidalequation}
\end{equation}
Here $v_k(r)(\equiv (GM/r)^{1/2})$ is the Keplerian speed at radius $r$.
We note that this equation describes the poloidal velocity of wind at radius $r$. On the one hand, the wind can be launched from any radius $\la r$.
On the other hand, we see in Figure \ref{Fig:massrate} that the wind mass flux increases rapidly with radius. Therefore, the mass-flux weighted wind velocity (\ref{poloidalequation}) primarily reflects wind launched
close to radius $r$.

\begin{figure}
\epsscale{1.1}\plotone{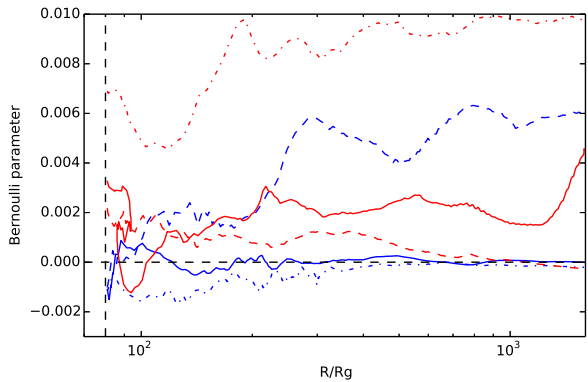}\epsscale{1.}
\caption{Same as the left panel of Figure \ref{Fig:poloidalspeed}, but for the Bernoulli parameter $Be$.}
\label{Fig:bernoulli}
\end{figure}

We have also calculated the evolution of the poloidal wind velocity along individual test particle trajectories.
The results are shown in Fig. \ref{Fig:poloidalspeed}. In the left panel, six representative test particles are
shown, initially located at $r=80r_g$ but different $\theta$. Specifically, the red dot-dashed line corresponds
to $\theta=10^{\circ}$, i.e, within the disk jet region; while others have $\theta\ga 20^{\circ}$. The three red
lines have a positive Bernoulli parameter at $80r_g$ while the three blue lines have $Be<0$ at $80r_g$
(refer to Fig. \ref{Fig:bernoulli}). The acceleration of the red dot-dashed line is the most significant,
indicating strong acceleration in the jet region. This is confirmed by our detailed analysis of the
acceleration mechanism in \S\ref{mechanism}. We see that the blue dashed and red solid lines in the left
panel of the figure also show strong acceleration. This is likely because these two test particles later enter
the ``jet region'' although they are initially located out of this region. In fact, we find that the $\theta$ values
of many test particles change significantly as they travel outward. For other lines (particles always in the
wind region), while there are fluctuations, the poloidal wind velocity roughly remain constant along the
particle trajectories, extending from $r=80r_g$ to $r\sim 800 r_g$, regardless of the sign of their initial $Be$. Beyond $r\sim 800r_g$, the poloidal velocity seems to decrease with radius. This is related to the value of $Be$. We will argue in the next subsection that such a decrease is likely not reliable.

In the right panel of Fig. \ref{Fig:poloidalspeed}, we show the averaged poloidal velocity of eight test particles
along their trajectories. The initial locations of these particles are uniformly distributed in $\theta$. The red and black lines correspond to
particles originated from $r=40 r_g$ and $160r_g$, respectively. The initial poloidal velocities of the red lines
are larger than that of the black lines, consistent with Fig. \ref{Fig:poloidalspeedradius}. We further distinguish
the dashed lines, which correspond to particles that are {\it initially} located in the wind region
($15^{\circ}\la \theta \la 165^{\circ}$), and the solid lines, corresponding to particles that {\it always} stay in
the wind region along their trajectories. We see that the particle poloidal velocity either keeps constant or
slightly increases outward along the trajectory. Solid and dashed lines differ significantly  because we are not using mass flux-weighted average. The increasing or constant behavior of the poloidal velocity with radius strongly suggests additional acceleration forces must
operate to compensate for the change in gravitational energy, which will be discussed in \S\ref{mechanism}.

From Fig. \ref{Fig:poloidalspeed}, we deduce that the asymptotic terminal poloidal wind velocity originated
from radius $r$ can be approximated by:
\begin{equation}
v_{\rm p,term}(r)\approx (0.2\sim 0.4)v_k(r).
\label{poloidalspeedtrajectory}
\end{equation}
Note the different meaning between this equation and eq. (\ref{poloidalequation}). This result also
compliments the discussion following eq. (\ref{poloidalequation}): the measured wind at any given
radius, say $r$, is a mixture of wind launched from smaller disk radii, and the wind that is produced
more locally. The former typically has higher velocity but carries smaller mass flux, while the latter
carries higher mass flux with smaller velocity. Overall, eqs. (\ref{poloidalequation}) and
(\ref{poloidalspeedtrajectory}) are approximately consistent with each other, and they are
also consistent with values estimated in Yuan, Bu \& Wu (2012; section 3.5).

\begin{figure*}
\epsscale{0.5}\plotone{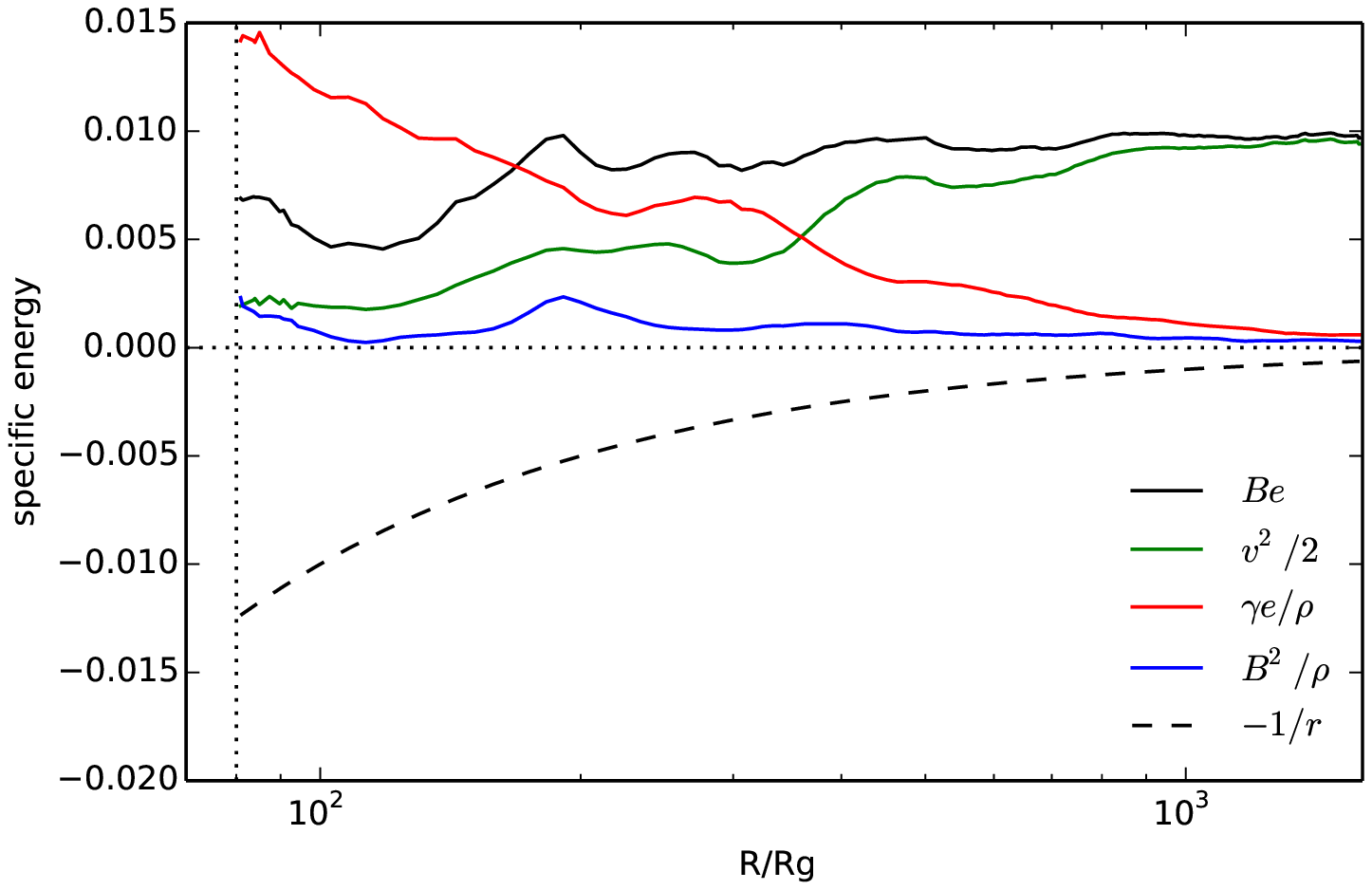}\epsscale{0.5}\plotone{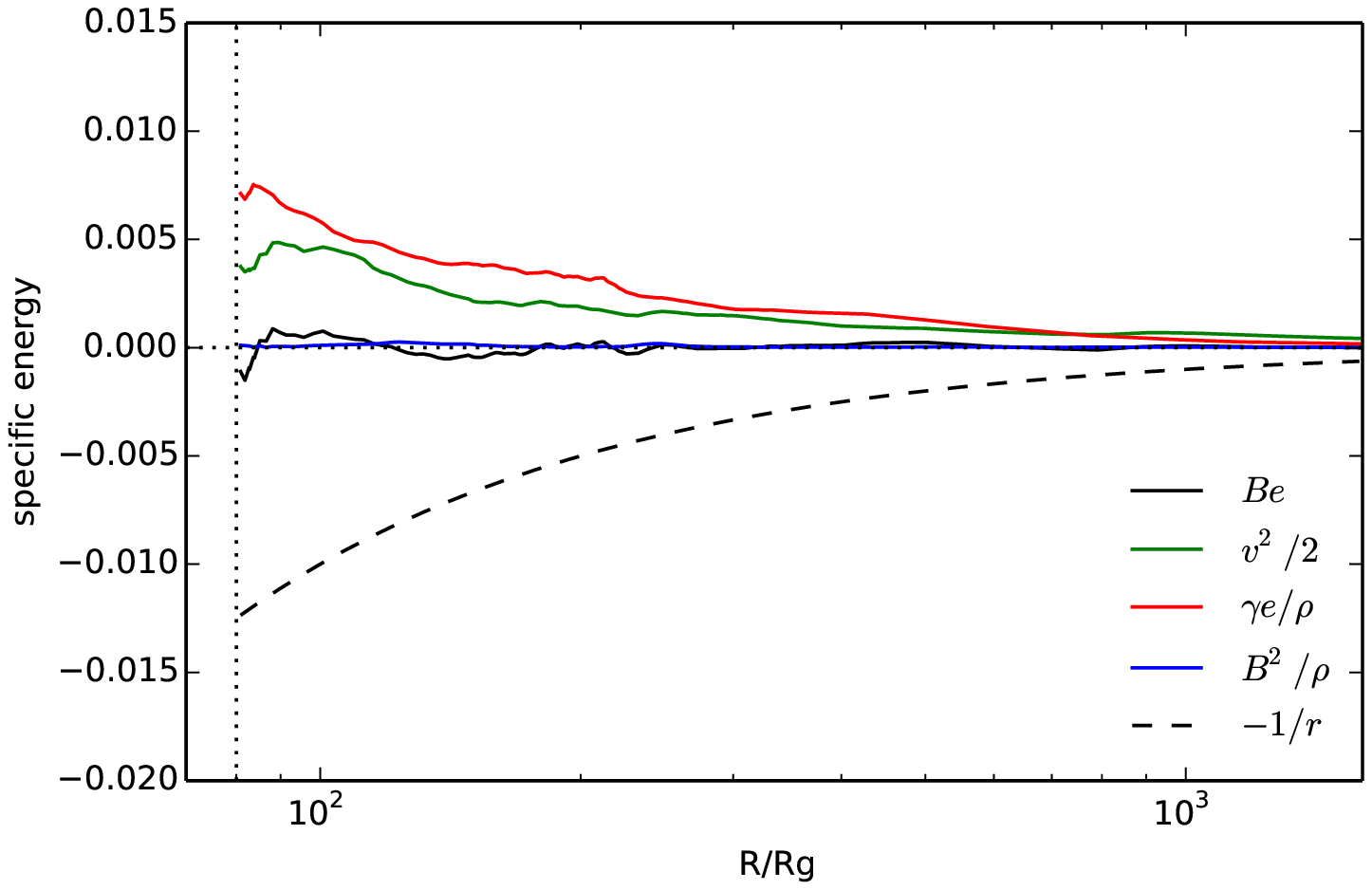}
\caption{Decomposition of the Bernoulli parameter (black solid) into individual contributions along
Lagrangian test particle trajectories for disk-jet ({\it left}) and wind ({\it right}). Individual terms include
gravitational energy ($\propto 1/r$; black dashed), specific kinetic energy ($\propto v^2/2$; green),
specific enthalpy ($\gamma e/\rho$; red), and specific magnetic energy ($B^2/\rho$; blue).
The test particles are initially located at $R=80r_g$, started at $t=100000M$.}
\label{Fig:bernoulliterms}
\end{figure*}

\subsection{The Bernoulli parameter of real outflow}
\label{bernoulli}

We now discuss the Bernoulli parameter of the wind.
Following Penna et al. (2013b), we adopt in the present paper the following definition of the Bernoulli
parameter with the magnetic term included,
\begin{equation}
Be=-\frac{\rho u_t+\Gamma u u_t+b^2u_t}{\rho}-1,
\end{equation}
where $\Gamma=5/3$ is the adiabatic index, $u_t$ is the time component of the four-velocity, $u$ is
the internal energy, $b$ is magnetic field strength in the fluid frame.
The rest mass energy is subtracted. Far from the black hole, the above equation reduces to the
Newtonian quantity, i.e., the sum of kinetic energy, gas enthalpy and magnetic enthalpy.

The evolution of the Bernoulli parameter along the trajectories of representative wind particles is
shown in Fig. \ref{Fig:bernoulli}. The three red lines have a positive $Be$ at the starting point
while the three blue lines have a negative initial value of $Be$. From this figure, we find the following
results.

First, the value of $Be$ is not a constant along particle trajectories.
This is not unexpected because conservation of $Be$ holds only when the accretion flow is strictly
steady and inviscid, while real accretion flow is always turbulent. As a result, it is inappropriate to use
the sign of $Be$ to judge whether the flow can escape to infinity, especially when the outflow is still within the radius at which turbulence is well developed.

Second, for outflow originated from smaller $\theta$, $Be$ increases to very large values at large radii.
On the other hand, the value of $Be$ varies much less significantly and roughly remains constant for
the wind originated from closer to the disk surface. Comparing with Fig. \ref{Fig:poloidalspeed}, we see
that changes in $Be$ correlates positively with changes in poloidal velocity. For the three trajectories
whose $Be$ increases outward, there is also significant acceleration in their poloidal velocities. This is
because that their $\theta$ values gradually decreases and enters into the jet region, thus experience strong acceleration.

It is interesting to note that for wind originated within $40^{\circ}\la\theta \la 50^{\circ}$, their value of
$Be$ becomes almost constant when $r\ga 800r_g$. This corresponds to the slight decrease of the poloidal velocity beyond $\sim 800r_g$ shown in the left panel of Fig. \ref{Fig:poloidalspeed}. The reason why $Be$ does not change beyond $800 r_g$ is because  in this region turbulence has not well developed within our simulation time. Note that this radius is different from the inflow equilibrium radius, which is $\sim 90r_g$. Within $\sim 90r_g$, everything, especially the radial density profile is fully reliable. Beyond this radius, the density profile is not reliable, but other properties, such as the level of turbulence and subsequently outflow properties, are still reliable up to a much larger ``turbulence radius'', the limiting radius of turbulence steady state. This radius can be estimated as follows. Turbulence in accretion flow is because of MRI. The fastest growth rate of MRI at radius $r$ is $\sim \Omega (r)$. More precisely, it takes $3\sim 4$ orbits for MRI to develop and $\sim 10$ orbits to saturate (Hawley et al. 1995). For our simulation time of $t_{\rm simulation}\sim 2\times 10^5$, taking a timescale of 3 orbits, we can obtain that the ``turbulence radius'' is $\sim 500 r_g$. This is close to the value of $800r_g$ mentioned above.  Another way to understand the ``turbulence radius'' is as follows. For a geometrically thick disk, the largest turbulence eddies have size of order $r$. The corresponding eddy turnover time is $r/\sigma(r)$, where $\sigma(r)$ is the {\it rms} turbulent velocity. Our simulation data shows $\sigma(r)\sim 0.15 v_k(r)$. If at a certain radius the eddy turnover time is substantially smaller than the duration of the simulation, then the local turbulence is likely to have reached quasi-steady state. Therefore the ``turbulence radius'' should be some fraction of $\sigma(r) t_{\rm simulation}$, which gives a similar result to the above estimation.
Therefore, we think that the results beyond $\sim 800r_g$ are not reliable. It is very likely that $Be$ will keep changing and the poloidal velocity still remains constant beyond $800 r_g$. This implies that wind can at least escape beyond the outer boundary of accretion flows. Simulations with longer run time can check this point.

Although $Be$ is not a constant along particle trajectories, it is still useful to decompose $Be$ into
individual physical terms and compare their contributions, as shown in Fig. \ref{Fig:bernoulliterms}.
The left and right panels correspond to particles originated from jet and wind regions, respectively.
In the case of disk jet, it is mainly the enthalpy that compensates for the increase of gravitational
energy and kinetic energy. Magnetic energy also plays an active role at smaller radius
($r\lesssim100r_g$). In the case of wind, the role of magnetic energy appears unimportant. The
increase of gravitational energy is mainly compensated by the reduction of specific enthalpy and
kinetic energy. But we note that although the total kinetic energy decreases along the trajectory,
the poloidal component does not. It usually keeps constant, as shown by Fig. \ref{Fig:poloidalspeed}.
This corresponds to the work done by the centrifugal force, as we will
discuss in \S\ref{mechanism}.

\subsection{The fluxes of energy and momentum  of wind and jet}

\begin{figure}
\epsscale{1.0}\plotone{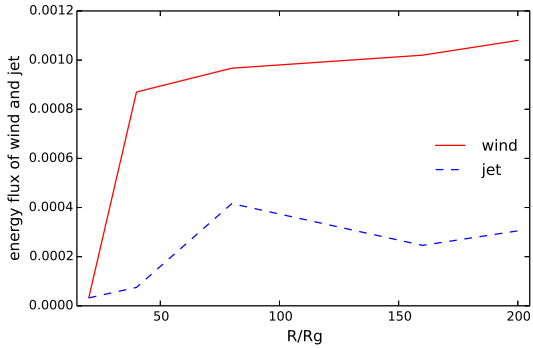}
\vspace{0.cm}
\caption{The radial profile of the energy fluxes of wind (red solid) and jet (blue dashed).}
\label{Fig:energyflux}
\end{figure}
\begin{figure}
\epsscale{1.0}\plotone{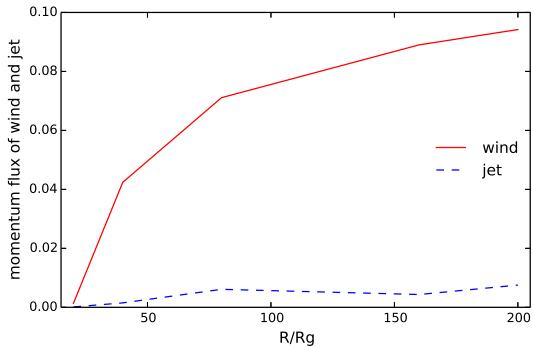}\vspace{0.cm}
\caption{The radial profile of the poloidal momentum fluxes of wind (red solid) and jet (blue dashed).}
\label{Fig:momentumflux}
\end{figure}

Based on the trajectory analysis, we now calculate the energy and momentum fluxes from both the wind and jet as follows.
 \begin{equation}
 \dot{E}_{\rm jet(wind)}(r)=\int \frac{1}{2}\rho(r,\theta,\phi)v_p^3(r,\theta,\phi)r^2\sin(\theta)d\theta d\phi,
 \end{equation}
 \begin{equation}
 \dot{P}_{\rm jet(wind)}(r)=\int \rho(r,\theta,\phi)v_p^2(r,\theta,\phi)r^2\sin(\theta)d\theta d\phi.
 \end{equation}
Here $v_p(r)$ is the poloidal velocity at radius $r$, which we assume $v_p(r)\approx v_r(r)$ in our calculation because we find $v_{\theta}\ll v_r$. The integration over $\theta$ for wind and disk-jet is bounded by $\theta\approx 15^{\circ}$, according to Fig. \ref{Fig:poloidalspeedtheta}.

Figs. \ref{Fig:energyflux} \& \ref{Fig:momentumflux} show the radial profiles of energy and momentum
fluxes, respectively, calculated at $T=100000M$. In both figures, the red solid and blue dashed lines are for the wind and disk jet,
respectively. We see that the energy flux of the wind is $\ga 3$ times stronger than that of the disk jet,
while the contrast in the momentum flux between wind and jet is much larger. This is again mainly because
of the low density in the disk-jet. From Fig. \ref{Fig:energyflux}, we see that the energy flux of the wind
rapidly increases at small radii, and then becomes almost saturated at $r\ga 40 r_g$. The rapid increase corresponds to the rapid increase of the wind mass flux with radius when $r\la 40 r_g$ (see Fig. \ref{Fig:massrate}). For $r\ga 40 r_g$, we have
\begin{equation}
\dot{E}_{\rm wind}(r)\approx  \frac{1}{2}\dot{M}_{\rm wind}(r) v^2_{\rm p,wind} (r)\approx \frac{1}{1000}\dot{M}_{\rm BH} c^2.
\label{energyflux}
\end{equation}
In the above calculation, eqs. (\ref{windflux}) and (\ref{poloidalequation}) are used.
This result indicates that the energy flux at large radius is roughly saturated, consistent with
Fig. \ref{Fig:energyflux}. The main reason energy flux saturates is $s=1$ in Eq. 5.
For the momentum flux of the wind, we have
\begin{equation}
\dot{P}_{\rm wind}(r) \approx \dot{M}_{\rm wind} (r)v_{\rm p,wind}(r)\propto r^{1/2}.
\label{momentumflux}
\end{equation}
This is consistent with the result shown in Fig. \ref{Fig:momentumflux}.

The energy flux obtained in eq. (\ref{energyflux}) is in good agreement with that required in large scale
AGN feedback simulations (e.g., Ciotti, Ostriker \& Proga 2010; Gaspari, Brighenti \& Temi 2012). In these
works, AGN feedback is involved to heat the inter-cluster medium to compensate for rapid cooling rate in
the systems (i.e., the cooling flow problem). It was found that to be consistent with observations of both
isolated galaxies and galaxy clusters, the required ``mechanical feedback efficiency", defined as
$\epsilon\equiv \dot{E}_{\rm wind}/\dot{M}_{\rm BH}c^2$, must be in the range of $\sim 10^{-3}-10^{-4}$.
Our results provide a natural explanation for the required value of $\epsilon$, at least when the AGN
is in the hot accretion mode.

Our results highlight the importance of wind over jet on AGN feedback. On the other hand, one important
caveat is that our calculation is based on the simulation of a Schwarzschild black hole. If the black hole is
rapidly spinning, the power of the disk jet is expected to become stronger (Sadowski et al. 2013).
In addition to the disk jet, a Poynting flux-dominated jet (BZ jet) will be produced through the black hole horizon,
powered by the black hole spin (Blandford \& Znajek 1976; Komissarov 2001; McKinney 2005;
Hawley \& Krolik 2006; Tchekhovskoy, Narayan \& McKinney 2010, 2011; Penna, Narayan \& Sadowski 2013a). The dependence of the power of BZ jet on spin is clear (see references above), while the dependence of the disk jet and wind on spin remains to be investigated, perhaps less sensitive compared to the BZ jet.

In addition to the black hole spin, another parameter is the magnetic flux threading the inner region of the accretion flow.  In our simulation, this flux is small and the accretion flow is called to be in the ``SANE'' state (Narayan et al. 2012). If the flux is large, the system enters the ``magnetically arrested disk'' (MAD) state.  There have been some studies on the dependence of the jet and wind power on the magnetic flux. For example, it has been found that in the MAD state, the power of the BZ jet will dominate the disk jet (Narayan, Igumenshchev \& Abramowicz 2003; Tchekhovskoy, Narayan \& McKinney 2010; Penna et al.2013a; see review by Yuan \& Narayan 2014). Sadowski et al. (2013) compared the power of the wind and the jet and found that, if the black hole spin and magnetic flux at the horizon are large, jet power will usually dominate the wind power. But note that as we have described in previous section, their estimation of the mass flux of
wind is should be regarded as a low limit. For a rapidly spinning black hole accreting in the  MAD limit, the power of the jet is even larger than the
accretion power (Tchekhovskoy, Narayan \& McKinney 2011; Tchekhovskoy, McKinney \& Narayan 2012). Systematic study on the dependence of the jet and wind power on magnetic flux and black hole spin is required. We plan to revisit  this problem in the next work.

\section{The mechanism of producing outflow}
\label{mechanism}

\subsection{Mechanism of acceleration of wind}

\begin{figure}
\epsscale{1.}\vspace{0.2cm}
\plotone{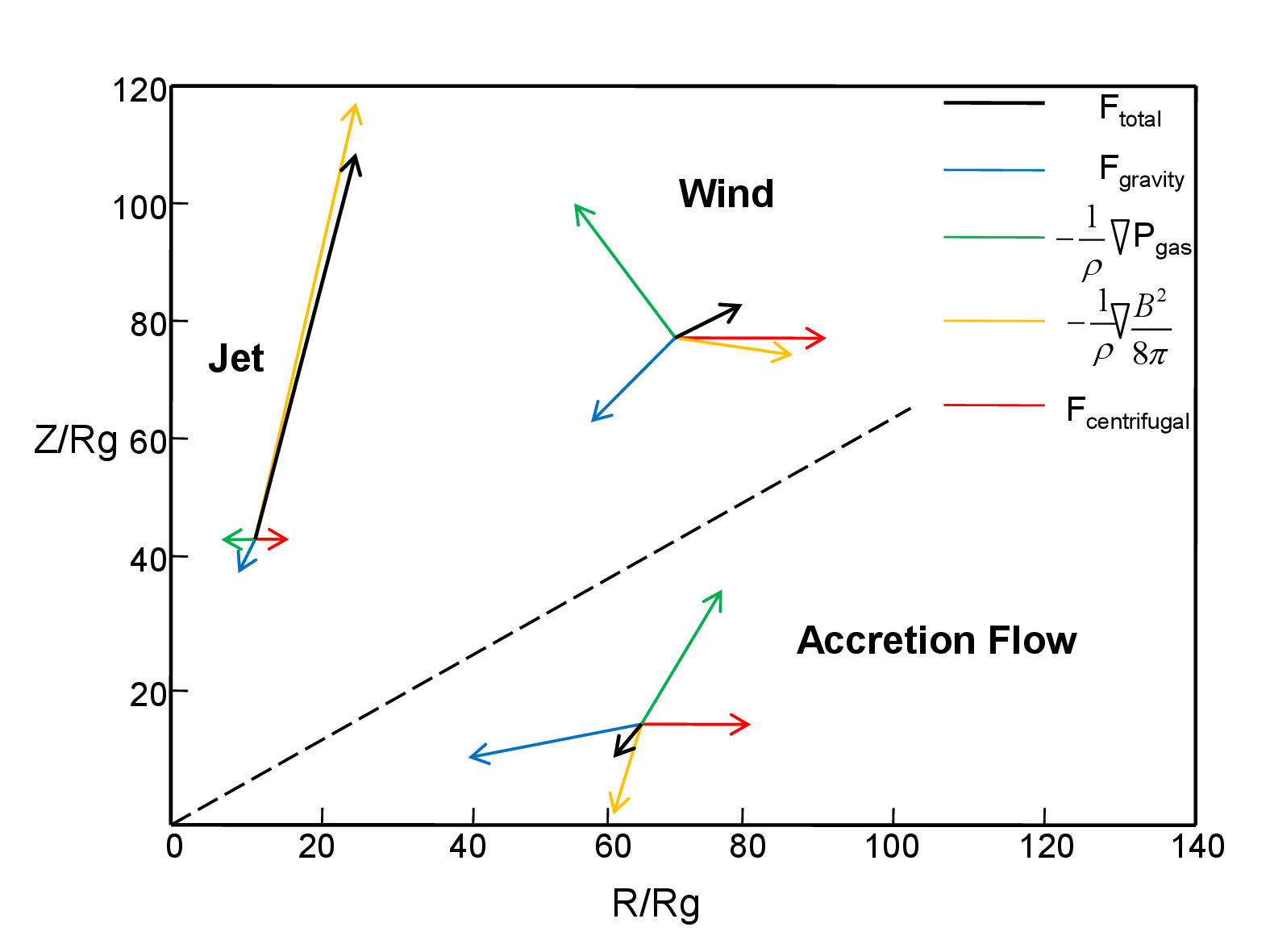}
\caption{Force analysis at three representative locations corresponding to the disk jet, wind and the main
body of the accretion disk. The arrows indicate force direction, whose length represents force magnitude.}
\label{Fig:force}
\end{figure}

To study the production and acceleration mechanism of wind and jet, we have calculated the forces
at the jet and wind region at a single point and a given snapshot. The locations where we evaluate the forces correspond to real outflow
based on our particle trajectory study. We note that forces are stochastic, but we found that our analysis can be regarded as representative, except in the cases we will mention below. Since we evaluate the forces in the co-rotating frame co-rotating with the flow at the ``evaluation location'', we should also include the centrifugal force, in addition to the gravitational force, gas pressure gradient, and the Lorentz
 force\footnote{We only include the gradient of magnetic pressure. Since $B_r\ll B_{\phi}$, the magnetic tension force is much weaker than the gradient of magnetic pressure so we neglect the tension force.}.
We show the results for three representative points in the disk jet, wind and main disk regions
in Figure \ref{Fig:force}. For the wind, the main driving forces are the centrifugal force and magnetic
pressure gradient. From the figure we notice that the gradient of the magnetic pressure is ``downward'',
pointing toward the positive $\theta$ direction. This is somewhat surprising because we usually expect
that magnetic field becomes weaker away from the main disk body toward the coronal region. This
reflects the strong fluctuation of the accretion flow. In fact, if we choose another time or another location
to do the force analysis, we very likely find that the gradient of the magnetic pressure becomes ``upward''.
Because of the same reason, the direction of the gas pressure gradient also strongly fluctuates with time
and location. But statistically, the gradients of both the gas and magnetic pressure are pointing along the
positive $r$ direction thus are helpful to the acceleration of wind. Their magnitudes are also comparable
to the centrifugal force, as shown by Fig. \ref{Fig:force}.

From the figure we can see that the magnitude of the centrifugal force is larger than (or at least
comparable to in general) the gravitational force. This means that the specific angular momentum of
the wind is larger than or close to the Keplerian value.
For comparison, we also show in the figure the force analysis in the inflow region. We see that the
magnitude of the centrifugal force is now smaller than the gravitational force\footnote{This confirms that the accretion flow is sub-Keplerian.}.
This is consistent with the result obtained in Yuan, Bu \& Wu (2012) that the specific angular momentum
of outflow is systematically larger than that of the inflow. This implies that some angular momentum is
transferred  from some fluid element to other, likely by the magnetic field lines from the main disk body to the coronal region. Once the combination of the centrifugal force and the pressure gradient
exceeds the gravitational force, wind will be accelerated.


According to the above analysis, the mechanism of the acceleration of wind is similar to the Blandofrd \& Payne (1982) mechanism in the sense that the centrifugal force plays an important role. The differences are that the gradient of the pressure plays a comparable role compared with the centrifugal force. In addition, there is no large-scale magnetic field formed in our simulation and the wind region is not force free.

\subsection{Mechanism of the acceleration of jet}

In the case of disk jet, the acceleration mechanism is completely different from the Blandford \& Payne
(1982). Here the dominant force is the gradient of the toroidal magnetic pressure, consistent with the result
of Hawley \& Krolik (2006). This mechanism is the so-called magnetic tower mechanism
(Lynden-Bell 2003; see also Shibata \& Uchida 1985; Kato, Mineshige \& Shibata 2004). The reason for
much higher jet velocity compared with the wind velocity is related with the strong magnetic field and low density.
In both the case of wind and jet, the energy of the outflow mainly comes from the rotation energy of the accretion flow. The rotation energy is converted into the magnetic energy which then is converted into the kinetic energy of the wind and jet.

\section{Summary}

Previous numerical simulations to the black hole accretion flow have shown that the mass accretion rate decreases inward (eq. (\ref{inflowratepowerlaw})). It has been proposed that  convection or outflow may result in this result. Stability analysis, however, has indicated that the accretion flow is convectively stable when the magnetic field is present. This excludes the possibility of convective instability and leave only the mass loss via outflow as the likely reason. However, previous theoretical works have obtained quite different result on the strength of outflow. In terms of the mass accretion rate onto the black hole horizon $\dot{M}_{\rm BH}$, Narayan et al. (2012) found that even at $r\approx 90r_g$, the mass flux of real outflow is still very weak. On the other hand, Yuan, Bu \& Wu (2012) argue that mass flux of outflow should be strong, i.e., a significant fraction of that described by eq. (\ref{outflowrate}).

One of the main aims of the present work therefore is to investigate how strong the real outflow is. It is well known that accretion flow is turbulent, so there must be some gas moving outward in any snapshot as part of turbulent eddies. These ``tubulent outflow'' are not ``real outflow'' since they will turn back and join the accretion flow. The main difficulty of obtaining the mass flux of real outflow is therefore how to exclude the ``contamination'' of turbulent outflow in eq. (\ref{outflowrate}). For this aim, instead of eq. (\ref{outflowrate}), Narayan et al. (2012) move the $t\phi$ average inside the integral in eq. (\ref{outflowrate}).

We have adopted a different approach to investigate this problem. We use a ``trajectory approach'' to analyze the data of GRMHD numerical simulation of accretion flow. Different from the streamline analysis often adopted in accretion flow study, this approach can provide the trajectory of each ``virtual test particles'' in the accretion flow thus directly show whether the the flow is turbulent outflow or real outflow. The most important result of our analysis is that the mass flux of real outflow is found to be as high as $\dot{M}_{\rm wind}\approx \dot{M}_{\rm BH}(r/40r_g)$ (i.e., eq. (\ref{windflux}); refer to Fig. \ref{Fig:massrate}). In other words, the mass flux of real outflow is equal to $\dot{M}_{\rm BH}$ at $40r_g$. As comparison, the mass flux calculated by  eq. (\ref{outflowrate}) is equal to $\dot{M}_{\rm BH}$ at $30r_g$.  The reason why Narayan et al. (2012) found a much weaker  outflow  is that the real outflow is instantaneous. They wander around in 3D space, as shown by Fig. \ref{Fig:angularrate}, thus will be cancelled if the time-average is done first.

Several other important results are as follows.
\begin{itemize}
\item Most of the real outflow occur in the coronal region of the accretion flow. Within the disk body, it is mainly inflow (Fig. \ref{Fig:trajectory}).
\item There are two distinct types of real outflow. One is within the region of $\theta \la 15^{\circ}$ away from the axis, and another is outside of this region. In the former region, the poloidal speed of outflow is as high as $\sim (0.3-0.4)c$; while in the latter region, the speed is much lower, $\la 0.05c$ (Fig. \ref{Fig:poloidalspeedtheta}). We call the outflow in the first region ``disk jet'' while the outflow in the second region ``wind''. ``Disk jet'' is different from the Blandford-Znajek jet in several ways, as summarized in Yuan \& Narayan (2014). The most interesting point is that disk jet exists even though the black hole is nonrotating.
\item For a given test particle, the poloidal speed of disk jet is found to increase along their trajectory; while for wind, the poloidal speed almost remains constant (Fig. \ref{Fig:poloidalspeed}). This implies that wind can at least escape beyond the outer boundary of the accretion flows. On the other hand, the poloidal speed of disk jet and wind decreases with increasing radius where they are produced (refer to eq. \ref{poloidalspeedtrajectory}). This implies that the wind has a mixture of poloidal speed depending on its original launching radius. But the mass flux-weighted poloidal speed of wind as a function of radius can be described by eq. (\ref{poloidalequation}) (Fig. \ref{Fig:poloidalspeedradius}).
\item The value of Bernoulli parameter $Be$ of real outflow is not a constant along their trajectories (Fig. \ref{Fig:bernoulli}). The physical reason is that the accretion flow is not steady but turbulent. Because of this reason, the value of  $Be$ for a real outflow is not necessarily positive.
\item The poloidal speed of outflow does not decrease along the trajectory. This indicates that  there must be some acceleration forces. We have analyzed the data and found that in the case of disk jet, the dominant acceleration force is the gradient of magnetic pressure. While for the wind, the centrifugal force and the gradient of gas and magnetic pressure play comparable roles (Fig. \ref{Fig:force}).
\item We have also calculated the fluxes of energy and momentum of wind and jet (Figs. \ref{Fig:energyflux} and \ref{Fig:momentumflux}). Especially, the kinetic energy flux of wind is described by eq. (\ref{energyflux}). The implied efficiency of wind production is $\epsilon_{\rm wind}\equiv \dot{E}_{\rm wind}/\dot{M}_{\rm BH}c^2\approx 10^{-3}$, in good agreement with the value required in large scale AGN feedback simulations.

\end{itemize}

\section*{Acknowledgments}
FY thanks Jim Stone for the valuable discussions. FY, ZG, and DB were supported in part by the National Basic Research Program of China (973 Program, grant 2014CB845800), the Strategic Priority Research Program ``The Emergence of Cosmological Structures'' of CAS (grant XDB09000000), and NSF of China (grants 11103059,
11103061, 11121062, and 11133005). RN was supported in
part by NSF grant AST1312651 and NASA grant NNX14AB47G. AS acknowledges support for this work by NASA through Einstein Postdoctotral Fellowship number PF4-150126 awarded by the Chandra X-ray Center, which is operated by the Smithsonian Astrophysical
Observatory for NASA under contract NAS8-03060. XB is supported by NASA through Hubble Fellowship grant HST-HF2-51301.001-A awarded by the Space Telescope Science Institute, which is operated by the Association of Universities for Research in Astronomy, Inc., for NASA, under contract NAS 5-26555. The authors
acknowledge computational support from NSF via XSEDE resources (grant
TG- AST080026N), from NASA via the High-End Computing (HEC)
Program through the NASA Advanced Supercomputing (NAS) Division at
Ames Research Center, and from SHAO Super Computing Platform.



\end{document}